\documentclass[prc,twocolumn,showpacs,preprintnumbers,amsmath,amssymb]{revtex4-1}
\usepackage[dvips]{graphicx}
\usepackage{dcolumn}
\usepackage{bm}
\usepackage{color}

\begin{document}
\title{ Progress in Many Body Theory with the Equation of Motion method. Time dependent Density Matrix meets Self-Consistent RPA. Applications to solvable Models.}

\author{ Peter Schuck$^{1,2}$, Mitsuru Tohyama$^{3}$}

\affiliation{$^1$Institut de Physique Nucl$\acute{e}$aire, IN2P3-CNRS,
Universit$\acute{e}$ Paris-Sud, F-91406 Orsay Cedex, France}
\affiliation{$^2$Laboratoire de Physique et de Mod\'elisation des Milieux Condens\'es, CNRS and Universit\'e Joseph Fourier, 25 Av. des Martyrs, BP 166, F-38042 }
\affiliation{$^3$Kyorin University School of Medicine, Mitaka, Tokyo
  181-8611, Japan     }

\begin{abstract}
The Bogoliubov-Born-Green-Kirkwood-Yvon or Time-Dependent Density Matrix (TDDM) hierarchy of equations for higher density
matrices is truncated at the three body level in approximating the three
body correlation function by a quadratic form of two body ones, closing
the equations in this way. The procedure is discussed in detail and it is
shown in non-trivial model cases that the approximate inclusion of three
body correlation functions is very important to obtain precise results. A
small amplitude approximation of this time dependent nonlinear equation
for the two body correlation function is performed (STDDM*-b) and it is shown that
the one body sector of this generalised non-linear second RPA equation
is equivalent to the Self-Consistent RPA (SCRPA) approach which had been
derived
previously by different techniques. It is discussed in which way SCRPA
also contains the three body correlations. TDDM and SCRPA are tested
versus exactly solvable model cases.
\end{abstract}
\maketitle
\section{Introduction}

Many body theory is well defined at the lowest order, that is, at the mean field level. Practically in all domains of many body physics the same type of mean field equations are applied, even though in detail there may be quite important deviations. This concerns, for instance,  density functional theory, e.g., \`a la Kohn-Sham \cite{Perdew} where already important many body correlations are incorporated in an equation for the single particle density (matrix). 
The cases where, like in atomic physics, one can work with a one body theory built on a non-renormalised bare force as is the case with the original Hartree-Fock theory, are quite rare. 
In spite of the extraordinary success of these ``effective'' mean field approaches, in many cases, there is need to go beyond and treat two, three, .. body correlations explicitly. Unfortunately, so far, no well accepted universal method applicable in practically same way in all domains, analogous to mean field theory, does not exist for higher correlation functions. 
Rather the situation is such that the way how correlations are treated is tailored to the problem at hand. There exist the Brueckner Hartree-Fock (BHF)~\cite{BHF} approach with extensions to treat the difficult hard core problem of the force as , e.g., in nuclear physics or in liquid $^3$He; there is the Gutzwiller wave function \cite{Gutz} to deal with double occupancies in lattice models; there are extensions of RPA together with various forms of Time dependent density matrix (TDDM) theory to be dealt with again in this work. 
Coupled Cluster theory (CCT) has become quite in vogue in chemistry ~\cite{CCT},~\cite{Bishop}. An important branch of many body physics is, of course, represented by the Quantum Monte Carlo (QMC) approaches ~\cite{QMC}, ~\cite{Markus} and also the very successful Density Matrix Renormalisation Group Methods (DMRG) ~\cite{DMRG}, ~\cite{Schollwoeck}. The method of correlated basis functions ~\cite{Clark} is a further promising theory. The list could be extended with many examples more.\\

\noindent
In such a disperse situation, we find it promising to present in this work the merging of two types of many body approaches which evolved so far independently from one another. We, indeed, discovered, and this will be the main subject of this paper, that the recently proposed extension of TDDM 
where the Bogoliubov-Born-Green-Kirkwood-Yvon (BBGKY) hierarchy of coupled time dependent density matrices is truncated at the three body level, approximating the three body density matrix by a quadratic form of two body densities leading to a self consistent closed non-linear equation for the two body density matrix,  has a close relation to the so-called Self-Consistent RPA (SCRPA) existing in the literature under various forms since quite some time. Both TDDM and SCRPA have in recent years shown their high efficiency in applications to several non-trivial model cases as well as to a few more realistic cases ~\cite{NPA628, NPA512, Delion, Hirsch, Storo, Jemai, Jemai2} .\\

\noindent
In this paper we will demonstrate the non-trivial relation of these two many body theories which start from very different ends, lending more credit to their well-foundedness and their wide spread applicability in several branches of physics. Applications to several model cases will further elucidate the structure of the theory.\\

The paper is organised as follows. In Sect.II, we describe our new decoupling method of TDDM where we give an expression for the 3-body correlation function $C_3$ in terms of a quadratic form of the two body correlation functions $C_2$. In Sect.III, the small amplitude limit (STDDM-b and STDDM*-b) of the coupled equations for the one body density matrix and the 2-body correlation functions is derived. In Sect.IV, we study the relation between Self-Consistent RPA (SCRPA) and STDDM and in Sect. V 
we demonstrate that to good approximation the Coupled Cluster two-body subsystem approximation (SUB2) 
wave function is the ground state of SCRPA and, thus, to a certain extent also of STDDM-b and STDDM*-b. In Sect.VI, a short outline of how SCRPA is related to a many body Green's function approach is presented. In Sect.VII, we show results of applications to a couple of exactly solvable models where the performances of the various methods can be appreciated.

\vspace{1.5cm}

\noindent
\section{Extended time dependent density matrix (TDDM) method}

\subsection{General formalism}

We will base our considerations on the following second quantized Hamiltonien with two body interactions written in a single particle basis where the single particle part of the Hamiltonian is diagonal (think, e.g., of kinetic energy in plane wave basis, or harmonic oscillator basis if the system is in an external quadratic potential as is mostly the case for trapped cold atoms).

\begin{equation}
H = \sum_{\alpha} e_{\alpha}a^+_{\alpha}a_{\alpha} + \frac{1}{4}\sum_{\alpha \beta \gamma \delta}\bar v_{\alpha \beta \gamma \delta}
a^+_{\alpha}a^+_{\beta}a_{\delta}a_{\gamma}.
\label{H}
\end{equation}

\noindent
Here, 
the $e_{\alpha}$'s are the single particle energies figuring together with the 2-body interaction part where the antisymmetrized matrix element of the force is defined by $\bar v_{\alpha \beta \gamma \delta}= \langle \alpha \beta|v|\gamma \delta \rangle -\langle \alpha \beta|v|\delta \gamma \rangle $.\\

The BBGKY hierarchy for density matrices with their equation of motion (EOM) is well documented in the literature, see, e.g., \cite{Bonitz} and references in there. 
It is straightforward to write down the first two of these equations which involve the one, two, and three body density matrices

\begin{eqnarray}
i\dot \rho_{\alpha \alpha'}&=&(e_{\alpha} - e_{\alpha'})\rho_{\alpha \alpha'}
\nonumber \\
&+&\frac{1}{2}\sum_{\lambda_1\lambda_2\lambda_3}
[\bar v_{\alpha\lambda_1\lambda_2\lambda_3} \rho_{\lambda_2\lambda_3\alpha'\lambda_1}
\nonumber \\
&-&\rho_{\alpha\lambda_1\lambda_2\lambda_3}\bar v_{\lambda_2\lambda_3\alpha'\lambda_1}],
\label{EOM1}
\end{eqnarray}

\begin{eqnarray}
i\dot \rho_{\alpha \beta \alpha' \beta'} &=& (e_{\alpha} + e_{\beta}-e_{\alpha'}-e_{\beta'})\rho_{\alpha \beta \alpha' \beta'}
\nonumber \\
&+&\frac{1}{2}\sum_{\lambda_1\lambda_2}
[\bar v_{\alpha\beta\lambda_1\lambda_2}{\rho}_{\lambda_1\lambda_2\alpha'\beta'}
-\bar v_{\lambda_1\lambda_2\alpha'\beta'}{\rho}_{\alpha\beta\lambda_1\lambda_2}]
\nonumber \\ 
&+&\frac{1}{2}\sum_{\lambda_1\lambda_2\lambda_3}
[\bar v_{\alpha\lambda_1\lambda_2\lambda_3} \rho_{\lambda_2\lambda_3\beta\alpha'\lambda_1\beta'}
\nonumber \\
&+&\bar v_{\lambda_1\beta\lambda_2\lambda_3} \rho_{\lambda_2\lambda_3\alpha\alpha'\lambda_1\beta'}
\nonumber \\
&-&\bar v_{\lambda_1\lambda_2\alpha'\lambda_3} \rho_{\alpha\lambda_3\beta\lambda_1\lambda_2\beta'}
\nonumber \\
&-&\bar v_{\lambda_1\lambda_2\lambda_3\beta'} \rho_{\alpha\lambda_3\beta\lambda_1\lambda_2\alpha'}],
\label{EOM2}
\end{eqnarray}
where $\rho_{\alpha \alpha'} =  \langle \Psi(t)|a^+_{\alpha'} a_{\alpha} |\Psi(t)\rangle$, $ \rho_{\alpha \beta ,\alpha' \beta'} = \langle \Psi(t)|a^+_{\alpha'} a^+_{\beta'}a_{\beta} a_{\alpha} |\Psi(t)\rangle$,  
$\rho_{\alpha \beta \gamma ,\alpha' \beta' \gamma'} = \langle \Psi(t)|a^+_{\alpha'} a^+_{\beta'}a^+_{\gamma'} a_{\gamma}a_{\beta} a_{\alpha} |\Psi(t)\rangle$ are the one, two, and three particle density matrices, respectively. 
The time dependent state is given by $|\Psi(t)\rangle = e^{-iHt}|\Psi(0)\rangle$. For a system consisting of two particles Eq. (\ref{EOM1}) and Eq. (\ref{EOM2}) without the three-body density matrix are exact.

It is preferable to introduce in (\ref{EOM1},\ref{EOM2}) instead of the two and three body density matrices their fully correlated counterparts $C_2$ and $C_3$

\begin{equation}
\rho_{\alpha \beta \alpha' \beta'} = {\mathcal A}(\rho_{\alpha \alpha'}\rho_{\beta \beta'}) + C_{\alpha \beta \alpha' \beta'}
\label{2-body}
\end{equation}

\begin{eqnarray}
\rho_{\alpha \beta \gamma , \alpha' \beta' \gamma'}&=& {\mathcal AS}(\rho_{\alpha \alpha'}\rho_{\beta \beta'}\rho_{\gamma \gamma'} +\rho_{\alpha \alpha'}C_{\beta \gamma \beta' \gamma'}) 
\nonumber \\
&+& C_{\alpha \beta \gamma , \alpha' \beta' \gamma'}
\label{3-body}
\end{eqnarray}

\noindent
where ${\mathcal A}$ and ${\mathcal S}$ shall indicate that the products in parentheses are properly antisymmetrised and symmetrised, respectively. 

The resulting equations can be found, e.g., in \cite{TS14}. For completeness, we will present them here again.

\begin{eqnarray}
i\dot{\rho}_{\alpha\alpha'}&=&
\sum_{\lambda}(\epsilon_{\alpha\lambda}{\rho}_{\lambda\alpha'}-{\rho}_{\alpha\lambda}\epsilon_{\lambda\alpha'})
\nonumber \\
&+&\frac{1}{2}\sum_{\lambda_1\lambda_2\lambda_3}
[\bar v_{\alpha\lambda_1\lambda_2\lambda_3} C_{\lambda_2\lambda_3\alpha'\lambda_1}
\nonumber \\
&-&C_{\alpha\lambda_1\lambda_2\lambda_3}\bar v_{\lambda_2\lambda_3\alpha'\lambda_1}],
\label{n}
\end{eqnarray}
\begin{eqnarray}
i\dot{C}_{\alpha\beta\alpha'\beta'}&=&
\sum_{\lambda}(\epsilon_{\alpha\lambda}{C}_{\lambda\beta\alpha'\beta'}
+\epsilon_{\beta\lambda}{C}_{\alpha\lambda\alpha'\beta'}
\nonumber \\
&-&\epsilon_{\lambda\alpha'}{C}_{\alpha\beta\lambda\beta'}
-\epsilon_{\lambda\beta'}{C}_{\alpha\beta\alpha'\lambda})
\nonumber \\
&+&B^0_{\alpha\beta\alpha'\beta'}+P^0_{\alpha\beta\alpha'\beta'}+H^0_{\alpha\beta\alpha'\beta'}
\nonumber \\
&+&\frac{1}{2}\sum_{\lambda_1\lambda_2\lambda_3}
[\bar v_{\alpha\lambda_1\lambda_2\lambda_3} C_{\lambda_2\lambda_3\beta\alpha'\lambda_1\beta'}
\nonumber \\
&+&\bar v_{\lambda_1\beta\lambda_2\lambda_3} C_{\lambda_2\lambda_3\alpha\alpha'\lambda_1\beta'}
\nonumber \\
&-&\bar v_{\lambda_1\lambda_2\alpha'\lambda_3} C_{\alpha\lambda_3\beta\lambda_1\lambda_2\beta'}
\nonumber \\
&-&\bar v_{\lambda_1\lambda_2\lambda_3\beta'} C_{\alpha\lambda_3\beta\lambda_1\lambda_2\alpha'}],
\label{N3C2}
\end{eqnarray}
where $C_{\alpha\beta\gamma\alpha'\beta'\gamma'}$ is the correlated part of the three-body density-matrix in (\ref{3-body})
which is neglected in the original version of TDDM \cite{WC}. 
The energy (mean field) matrix $\epsilon_{\alpha\alpha'}$ is given by
\begin{eqnarray}
\epsilon_{\alpha\alpha'}=e_\alpha\delta_{\alpha\alpha'}
+\sum_{\lambda_1\lambda_2}
\bar v_{\alpha\lambda_1\alpha'\lambda_2} 
\rho_{\lambda_2\lambda_1}
\label{hf}.
\end{eqnarray}
The matrix $B^0_{\alpha\beta\alpha'\beta'}$ in Eq. (\ref{N3C2}) does not contain $C_{\alpha\beta\alpha'\beta'}$ and 
describes the $2p-2h$ and $2h-2p$ excitations:
\begin{eqnarray}
B^0_{\alpha\beta\alpha'\beta'}&=&\sum_{\lambda_1\lambda_2\lambda_3\lambda_4}
\bar v_{\lambda_1\lambda_2\lambda_3\lambda_4}
\nonumber \\ 
&\times&[(\delta_{\alpha\lambda_1}-\rho_{\alpha\lambda_1})(\delta_{\beta\lambda_2}-\rho_{\beta\lambda_2})
\rho_{\lambda_3\alpha'}\rho_{\lambda_4\beta'}
\nonumber \\
&-&\rho_{\alpha\lambda_1}\rho_{\beta\lambda_2}(\delta_{\lambda_3\alpha'}-\rho_{\lambda_3\alpha'})
(\delta_{\lambda_4\beta'}-\rho_{\lambda_4\beta'})].
\nonumber \\
\label{B^0}
\end{eqnarray}
Particle - particle and $h-h$ correlations 
are described by $P^0_{\alpha\beta\alpha'\beta'}$
\begin{eqnarray}
P^0_{\alpha\beta\alpha'\beta'}&=&\frac{1}{2}\sum_{\lambda_1\lambda_2\lambda_3\lambda_4}
\bar v_{\lambda_1\lambda_2\lambda_3\lambda_4}
\nonumber \\ 
&\times&[(\delta_{\alpha\lambda_1}\delta_{\beta\lambda_2}
-\delta_{\alpha\lambda_1}\rho_{\beta\lambda_2}
-\rho_{\alpha\lambda_1}\delta_{\beta\lambda_2})
{C}_{\lambda_3\lambda_4\alpha'\beta'}
\nonumber \\
&-&(\delta_{\lambda_3\alpha'}\delta_{\lambda_4\beta'}
-\delta_{\lambda_3\alpha'}\rho_{\lambda_4\beta'}
-\rho_{\lambda_3\alpha'}\delta_{\lambda_4\beta'})
{C}_{\alpha\beta\lambda_1\lambda_2}].
\nonumber \\
\label{P^0}
\end{eqnarray}
$H^0_{\alpha\beta\alpha'\beta'}$ contains the  $p-h$ correlations.
\begin{eqnarray}
H^0_{\alpha\beta\alpha'\beta'}&=&\sum_{\lambda_1\lambda_2\lambda_3\lambda_4}
\bar v_{\lambda_1\lambda_2\lambda_3\lambda_4}
\nonumber \\ 
&\times&[\delta_{\alpha\lambda_1}(\rho_{\lambda_3\alpha'}{C}_{\lambda_4\beta\lambda_2\beta'}
-\rho_{\lambda_3\beta'}{C}_{\lambda_4\beta\lambda_2\alpha'})
\nonumber \\
&+&\delta_{\beta\lambda_2}(\rho_{\lambda_4\beta'}{C}_{\lambda_3\alpha\lambda_1\alpha'}
-\rho_{\lambda_4\alpha'}{C}_{\lambda_3\alpha\lambda_1\beta'})
\nonumber \\
&-&\delta_{\alpha'\lambda_3}(\rho_{\alpha\lambda_1}{C}_{\lambda_4\beta\lambda_2\beta'}
-\rho_{\beta\lambda_1}{C}_{\lambda_4\alpha\lambda_2\beta'})
\nonumber \\
&-&\delta_{\beta'\lambda_4}(\rho_{\beta\lambda_2}{C}_{\lambda_3\alpha\lambda_1\alpha'}
-n_{\alpha\lambda_2}{C}_{\lambda_3\beta\lambda_1\alpha'})].
\nonumber \\
\label{H^0}
\end{eqnarray}

 So far things have been straightforward. The difficulty lies in the fact that the hierarchy of equations has to be decoupled in order to be applicable and to yield a self-contained system of equations. Many decoupling schemes have been proposed in the past, see, e.g., \cite{Bonitz}. 
 In nuclear physics the decoupling scheme of Cassing and Wang is often applied. It consists in neglecting the 3-body correlation function in (\ref{N3C2}) ($C_3$) altogether \cite{WC}. This then leads to a closed system of equations where the two body correlation matrix ($C_2$) figures linearly. 
 Recently the present authors have shown that three body correlations are very important \cite{TS10}.  Since they are difficult to incorporate fully, it was proposed {\it not to skip} $C_3$ entirely but to replace it by a quadratic form in $C_2$ \cite{TS14}. This then yields a closed system of non-linear equations for the two body correlation functions $C_2$. This approximation scheme is explained in the next section and in App. A.

\noindent
\subsection{Quadradic form of $C_3$ in terms of $C_2$'s}

Roughly speaking, the way how to express the three body correlation functions, $C_3$, as a quadratic form of two body ones ($C_2$) goes as follows. 
It is well known that the in medium 3-body Green's function can be expanded in analogy to the free three body problem into a series of 2-body in medium $T$-matrices, the equivalent of our $C_2$ correlation functions \cite{Faddeev}.

\begin{figure}
\includegraphics[width=8.6cm]{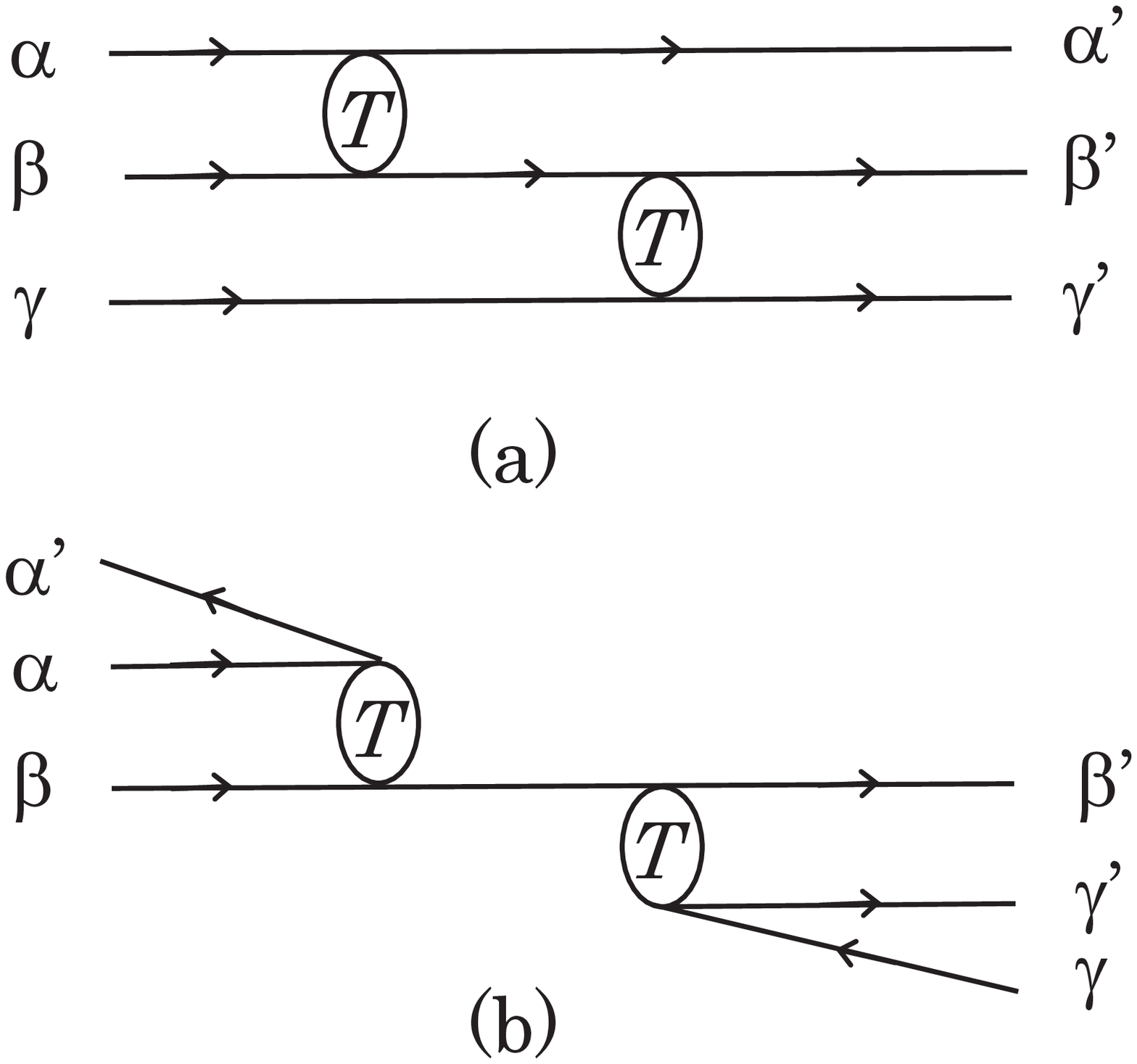}
\caption{\label{fig.1-2} One line reducible contributions to the 2$p$-1$h$(2$h$-1$p$) propagator. $T$ denotes the in-medium  2-body $T$ matrix.}
\end{figure}

In Fig.\ref{fig.1-2} we show graphically the second order contribution to the 3-body propagator in terms of the in-medium $T$-matrices. 
The first order terms are the ones Cassing and Wang have considered. We want to keep also the second order terms. 
Imposing a specific time ordering where two particles and one hole (2$p$-1$h$) (or 2$h$-1$p$) are traveling together as it may be deduced from the 3-body terms in (\ref{N3C2}), we obtain for the second order terms the second graph in Fig.\ref{fig.1-2}. 
One easily checks that there are 9 combinations of this type possible. 
Such contributions are contained in the 2$p$-1$h$ (2$h$-1$p$) propagator and are known as their one line reducible part, since they can be separated into two pieces in just cutting one line. 
The remainder is the so-called one line irreducible part and enters, e.g., the dynamic part of the single particle self energy in Dyson's equation for the single particle propagator \cite{RS}. 
The fact that an important part of the 2$p$-1$h$ (2$h$-1$p$) many body propagator can be separated into some quadratic form of two body propagators is already manifest at this point. 
Let us, however, be more analytic, since this will be a basic aspect of our theory. 
In \cite{RS}, App. F, it is shown that the one line reducible part of the $2p-1h (2h-1p)$ propagator can be expressed in a way which is shown in App. A.1 and graphically interpreted in Fig.\ref{fig.1-2}. A more direct but less intuitive way is obtained using identity relations of density matrices via their trace relations, see App. A.2. We give here the final result on which our approach will be based. Let us consider the first 3-body term in  Eq. (\ref{N3C2})    that is
 the three-body correlation matrix $C_{\lambda_2\lambda_3\beta\alpha'\beta'\lambda_1}$
 is written as
\begin{eqnarray}
C_{\lambda_2\lambda_3\beta\alpha'\beta'\lambda_1}
&=&\frac{1}{3-n_{\lambda_2}-n_{\lambda_3}-n_{\beta}-n_{\alpha'}-n_{\beta'}-n_{\lambda_1}}
\nonumber \\
&\times&\sum_\gamma(
-C_{\lambda_2\lambda_3\alpha'\gamma}C_{\beta\gamma\beta'\lambda_1}
-C_{\lambda_2\lambda_3\lambda_1\gamma}C_{\beta\gamma\alpha'\beta'}
\nonumber \\
&-&C_{\lambda_2\lambda_3\gamma\beta'}C_{\beta\gamma\alpha'\lambda_1}
-C_{\lambda_2\beta\alpha'\gamma}C_{\gamma\lambda_3\beta'\lambda_1}
\nonumber \\
&-&C_{\lambda_2\beta\beta'\gamma}C_{\lambda_3\gamma\alpha'\lambda_1}
-C_{\lambda_2\beta\gamma\lambda_1}C_{\lambda_3\gamma\alpha'\beta'}
\nonumber \\
&-&C_{\lambda_2\gamma\alpha'\beta'}C_{\lambda_3\beta\lambda_1\gamma}
-C_{\lambda_2\gamma\alpha'\lambda_1}C_{\lambda_3\beta\gamma\beta'}
\nonumber \\
&-&C_{\lambda_2\gamma\beta'\lambda_1}C_{\lambda_3\beta\alpha'\gamma}).
\label{3body2}
\end{eqnarray}
where we supposed that the single density matrices are  diagonal, that is

\begin{equation}
\rho_{\alpha \alpha'} = n_{\alpha}\delta_{\alpha \alpha'}
\label{occs}
\end{equation}
with $n_{\alpha}$ the single particle (s.p.)occupation numbers (one can always work in the basis where $\rho_{\alpha \alpha'}$ is diagonal 
but, below, we will argue that the non-diagonal terms of the s.p. density matrix are of higher order anyway).
As predicted from the graphical analysis, there are nine terms. They may not all be of same importance. 
On the other hand, the correlated part of the single particle occupations in the denominator may give raise to 
contributions which are of the same order of magnitude as the genuine four body correlations which have been neglected in (\ref{3body2}). 
So we will replace the occupation numbers by their mean field values 1 or 0. Before, we come to these further approximations, 
let us analyse the content  of the nine terms in (\ref{3body2}). Inserting the expression (\ref{3body2}) into (\ref{N3C2}) one may realise that this integral equation for the $C_2$'s couples all channels, that is the $C_2$'s in the quadratic terms are interconnected in all possible ways. This is reminiscent of what is done in parquet diagram technique, see, e.g., \cite{BR}. However, there is an important difference: in parquet diagrams the correlation functions are dynamic ones depending in general on three energies whereas here the correlation functions only depend on one energy which corresponds to the channel considered from the outset.

Above equations constitute our most general non-linear set for the 
calculation of the two body correlation function $C_2$. 

\noindent
\subsection{Static limit. Restriction to particle and hole indices}
Let us write down (\ref{N3C2}) for the static case

\begin{eqnarray}
0&=&
(\epsilon_{\alpha}
+\epsilon_{\beta}
-\epsilon_{\alpha'}
-\epsilon_{\beta'})C_{\alpha\beta\alpha'\beta'}
+B^0_{\alpha\beta\alpha'\beta'}+P^0_{\alpha\beta\alpha'\beta'}\nonumber\\
&+&H^0_{\alpha\beta\alpha'\beta'}
  + T_{\alpha\beta\alpha'\beta'}  
\label{static}
\end{eqnarray}

\noindent
where we supposed to work in a single particle basis where the single particle energies (\ref{hf}) are diagonal. 
As above, the three body part is given by

\begin{eqnarray}
&T&_{\alpha\beta\alpha'\beta'} = \nonumber\\    
&\frac{1}{2}&\sum_{\lambda_1\lambda_2\lambda_3}
[\bar v_{\alpha\lambda_1\lambda_2\lambda_3} C_{\lambda_2\lambda_3\beta\alpha'\lambda_1\beta'}
+\bar v_{\lambda_1\beta\lambda_2\lambda_3} C_{\lambda_2\lambda_3\alpha\alpha'\lambda_1\beta'}
\nonumber \\
&-&\bar v_{\lambda_1\lambda_2\alpha'\lambda_3} C_{\alpha\lambda_3\beta\lambda_1\lambda_2\beta'}
-\bar v_{\lambda_1\lambda_2\lambda_3\beta'} C_{\alpha\lambda_3\beta\lambda_1\lambda_2\alpha'}].
\label{three-body}
\end{eqnarray}

In general, we will not consider the 2-body correlation functions with arbitrary indices. Since, with particle ($p$) indices above and hole ($h$) indices below the Fermi level, the inhomogeneous term $B^0$, 
in the uncorrelated limit, is only non-zero for $C_{p_1p_2h_1h_2}$ and $C_{h_1h_2p_1p_2}= C^*_{p_1p_2h_1h_2}$, they are dominant but we consider additionally the following three index-combinations
\[C_{p_1h_1p_2h_2}; C_{p_1p_2p_3p_4}; C_{h_1h_2h_3h_4}, \] 
because they couple each other. One can suppose that they give the dominant contributions. One easily imagines that $C_2$'s with an odd number of either $p$- or $h$-indices are suppressed with respect to the ones with an even number of $p$($h$) indices.

In the past the  three body term $T$ was usually neglected \cite{WC}. Here we want to treat it in the approximate form given above. We have  four index combinations of $T$. In addition 
at least for situations not close to a macroscopic phase transition or to systems with a Goldstone (zero) mode, the single following term out of the nine possible is dominant

\begin{equation}
C_{p_1h_1h_2,p_2h_3h_4} \simeq \sum_pC_{p_1ph_3h_4}C_{h_1h_2p_2p}
\label{C_2C_2-h}
\end{equation}
that is the product of two correlation functions with 2p-2h indices is the most important one. There exists only one further 3-body correlation function which has this specific product property

\begin{equation}
C_{h_1p_1p_2,h_2p_3p_4} \simeq \sum_h C_{p_1p_2h_2h}C_{hh_1p_3p_4}.
\label{C_2C_2-p}
\end{equation}

Respecting this approximation, we obtain for the four possible three body terms

\begin{eqnarray}
&T&_{p_1p_2h_1h_2}=\frac{1}{2}\sum_{pp'hh'}
[\bar v_{ p_1phh'} C_{p_2p'h_1h_2}C_{hh'pp'} - ( p_1 \leftrightarrow p_2)]\nonumber\\
&+&\frac{1}{2}\sum_{hh'pp'}[\bar v_{pp'h_1h}C_{p_1p_2h_2h'}C_{h'hpp'} - (h_1 \leftrightarrow h_2)],
\label{T2p2h}
\end{eqnarray}
\begin{eqnarray}
T_{p_1h_1p_2h_2}&=&\frac{1}{2}\sum_{pp'hh'}\bar v_{ p_1hh'p} C_{p'ph_2h}
C_{h_1h'p_2p'}
\nonumber \\
&+&\frac{1}{2}\sum_{pp'p''h}\bar v_{ p_1pp'p''} C_{p'p''hh_2}
C_{h_1hp_2p}
\nonumber \\
&-&\frac{1}{2}\sum_{pp'hh'}\bar v_{ ph'p_2h} C_{hh_1pp'}
C_{p_1p'h'h_2}
\nonumber \\
&-&\frac{1}{2}\sum_{pp'p''h}\bar v_{ p'p''p_2p} C_{p_1ph_2h}
C_{hh_1p'p''}
\nonumber \\
&-&\frac{1}{2}\sum_{pp'hh'}\bar v_{ h_1pp'h} C_{h'hp_2p}
C_{p'p_1h'h_2}
\nonumber \\
&+&\frac{1}{2}\sum_{phh'h''}\bar v_{h_1hh'h''} C_{h'h''p_2p}
C_{pp_1hh_2}
\nonumber \\
&+&\frac{1}{2}\sum_{pp'hh'}\bar v_{ph'h_2p'} C_{p_1p'hh'}
C_{h_1hp_2p}
\nonumber \\
&-&\frac{1}{2}\sum_{phh'h''}\bar v_{h'h''hh_2} C_{pp_1h'h''}
C_{h_1hp_2p}.
\label{Tph}
\end{eqnarray}
The exchange matrix $T_{h_1p_1p_2h_2}$ of $T_{p_1h_1p_2h_2}$ is given not by changing $p_1$ and $h_1$ on the right-hand side of 
 Eq. (\ref{Tph}) but by using Eqs. (\ref{three-body}), (\ref{C_2C_2-h}) and (\ref{C_2C_2-p}). Then the exchange property $T_{h_1p_1p_2h_2}=-T_{p_1h_1p_2h_2}$
is satisfied. Furthermore, we have

\begin{eqnarray}
&T&_{p_1p_2p_3p_4} = \frac{1}{2}\sum_{phh'h''}[\bar v_{p_1hph'}C_{pp_2hh''}C_{h''h'p_3p_4} - ( p_1 \leftrightarrow p_2)]
\nonumber \\
&-& \frac{1}{2}\sum_{phh'h''}[\bar v_{ph'p_3h}C_{p_1p_2hh''}C_{h''h'pp_4} - ( p_3 \leftrightarrow p_4)].
\label{T4p}
\end{eqnarray}
\begin{eqnarray}
&T&_{h_1h_2h_3h_4}=\frac{1}{2} \sum_{pp'p''h} [\bar v_{h_1pp'h}C_{hh_2p''p}C_{p'p''h_3h_4}  - ( h_1 \leftrightarrow h_2)]
\nonumber \\
&-&\frac{1}{2} \sum_{pp'p''h} [\bar v_{p'hh_3p}C_{h_1h_2p''p'}C_{pp''hh_4}  - ( h_3 \leftrightarrow h_4)].
\label{T4h}
\end{eqnarray}

Inserting these four three body terms into (\ref{static}), one obtains a set of four coupled equations for the four possible two body correlation functions. It is this set of equations which will be used in the TDDM applications presented below. In principle it is straightforward to include into the three body terms all nine quadratic forms deduced with their specific combinations of particle and hole indices from (\ref{3body2}). However, this leads to a much more extended set of equations. In the numerical examples treated below, this does not seem necessary. However, as already mentioned, there may exist situations where the full set of equations is needed.

 Above four coupled equations for the four different $C_2$'s have a number of appealing properties. They are totally antisymmetric and they are number and energy conserving. The latter properties can easily be verified in just taking into account the (anti)symmetry properties of the equations. Other properties will be discussed in Sect.IV.B. To obtain the exact solution for a general 2-body problem, we have to discard the three body density matrix in (\ref{EOM2}). So, our equations which approximately include 3-body correlations are only valid for particle numbers $N \ge 3$.

\subsection{Procedures to obtain a static solution}

Two methods have been used to obtain a stationary state (ground state) of Eqs. (\ref{n}) and (\ref{N3C2}). One is 
the gradient method \cite{TTS04,Takahara} and the other a time-dependent method . Since the latter is used in the applications of TDDM
to the model Hamiltonians below, we explain it in some detail. The starting point is a non-interacting  ground state where the single-particle states below the Fermi level are completely occupied. 
(In the application to $^{16}$O \cite{Toh15}, we have also used the HF ground state 
where the mean-field effect is already included.) Then we solve Eqs. (\ref{n}) and (\ref{N3C2}) by gradually increasing 
the strength of the interaction such that $v({\bf r},t)=v({\bf r})$$\times t/T$. To suppress mixing
of excited states, we must take sufficiently large $T$: For example, $T\gg 2\pi/E_{2p-2h}$ where $E_{2p-2h}$ is the excitation energy of
a 2p-2h state.
This method is motivated by
the Gell-mann-Low theorem \cite{gell} and has often been used to obtain nearly stationary solutions of time-dependent problems \cite{pfitz,lacroix}.

Let us explain how this method works using an illustrative case. We try to obtain a perturbative expression
for $n_p$ assuming that only $C_{pp'hh'}$ and $C_{hh'pp'}$ are important. Under this assumption
Eq. (\ref{n}) for $n_p$ is written as 
\begin{eqnarray}
i\dot{n}_p=\frac{1}{2}\sum_{hh'p'}[\bar{v}_{pp'hh'}C_{hh'pp'}-\bar{v}_{hh'pp'}C_{pp'hh'}].
\label{purtn}
\end{eqnarray}
If we keep only the $B^0_{pp'hh'}$ term in Eq. (\ref{N3C2}), the equation for $C_{pp'hh'}$ is
\begin{eqnarray}
i\dot{C}_{pp'hh'}=(\epsilon_p+\epsilon_{p'}-\epsilon_h-\epsilon_{h'})C_{pp'hh'}+\bar{v}_{pp'hh'}\frac{t}{T}.
\nonumber \\
\label{purtC}
\end{eqnarray}
Equation (\ref{purtC}) is solved as 
\begin{eqnarray}
{C}_{pp'hh'}=-i\bar{v}_{pp'hh'}\int_0^t\frac{t'}{T}
e^{iE_{pp'hh'}t'}dt'e^{-iE_{pp'hh'}t},
\nonumber \\
\label{purtC1}
\end{eqnarray}
where $E_{pp'hh'}=\epsilon_p+\epsilon_{p'}-\epsilon_h-\epsilon_{h'}$.
Under the assumption $E_{pp'hh'}T>E_{pp'hh'}t\gg 1$, Eq. (\ref{purtC1}) gives $C_{pp'hh'}\approx -\bar{v}_{pp'hh'}{t}/{T}E_{pp'hh'}$,
which is the perturbative expression for the two-body correlation matrix.
Inserting Eq. (\ref{purtC1}) into Eq. (\ref{purtn}) and assuming that $E_{pp'hh'}t\gg 1$, we arrive at
the perturbative expression for the occupation probability of an unoccupied state
\begin{eqnarray}
n_p\approx\frac{1}{2}\sum_{hh'p'}\frac{|\bar{v}_{pp'hh'})|^2}{E_{pp'hh'}^2}\left(\frac{t}{T}\right)^2.
\end{eqnarray}
 Having approximate expressions for very small times $t/T$ for $n_p$ and $C_{pp'hh'}$, we insert those into r.h.s. of Eqs. (\ref{n}) and (\ref{N3C2}), assuming that single particle density matrix is diagonal and make one further time integration, incrementing time by a small step. This creates on l.h.s. a s.p. density matrix and a new $C_2$. Repeating this procedure until $t=T$, one arrives at the stationary solution for $C_2$ and the single particle density matrix $\rho_{\alpha \beta}$. Most of the time, at the end of the calculation, one either works in the basis (the 'canonical' basis) which diagonalises the density matrix as in (\ref{occs}) or one supposes that the density matrix is approximately diagonal what generally is verified to good approximation.

\section{Small amplitude limit of TDDM (STDDM and STDDM*)}

\subsection{Derivation of STDDM with non-linear terms (STDDM*)}

It is well known that time dependent HF leads to standard RPA (with exchange) in the small amplitude limit \cite{RS}. So, HF is the consistent ground state when the two body correlations $C_2$ are neglected. On the contrary, considering in addition the inclusion of two body correlations, i.e., the coupled system of equations (\ref{n}) and (\ref{N3C2}), the corresponding ground state will contain correlations. It will be interesting to see, in how far we can give an explicit expression for this ground state and in which way standard RPA is modified due to the inclusion of ground state correlations. So, let us take the small amplitude limit of present form of TDDM, i.e.,  Eqs. (\ref{n}) and (\ref{N3C2})   . With

\[
\rho_1 = \rho^{(0)}_1 + \delta \rho_1;~~~~~ C_2 = C_2^{(0)} + \delta C_2
\]

\noindent
and

\[ \delta \rho_1 = \sum_{\nu}[ \tilde \chi^{\nu}e^{-i\Omega_{\nu}t} +\tilde \chi^{\nu,+}e^{i\Omega_{\nu}t}];\]
\[\delta C_2 = \sum_{\nu}[\tilde {\mathcal X}^{\nu}e^{-i\Omega_{\nu}t} + \tilde {\mathcal X}^{\nu,+}e^{i\Omega_{\nu}t}] \]

\noindent
we obtain coupled equations for the one-body and two-body transition amplitudes
$\tilde \chi^\nu_{\alpha\alpha'}=\langle \nu|a^+_{\alpha'}a_{\alpha}|0\rangle$ and $\tilde {\mathcal X}^\nu_{\alpha\beta\alpha'\beta'}=\langle \nu|:a^+_{\alpha'}a^+_{\beta'}a_{\beta}a_{\alpha}:|0\rangle$:

\begin{equation}
\begin{pmatrix}a&b\\c&\tilde d
\end{pmatrix}
\begin{pmatrix} \tilde \chi^\nu \\ \tilde {\mathcal X}^\nu \end{pmatrix}
= \Omega_\nu
\begin{pmatrix} \tilde \chi^\nu \\ \tilde {\mathcal X}^\nu \end{pmatrix}.
\label{STDDM0}
\end{equation}
The matrix $\tilde d$ is written as $\tilde d=d+\Delta d$ where $d$ stems from variation of the linear terms of the two-body correlation matrix whereas
$\Delta d$ comes from the variation of the three-body correlation matrix when it is approximated as, e.g., in Eq. (\ref{C_2C_2-h}) and (\ref{C_2C_2-p}) by quadratic forms of $C_2$'s (that is the leading contributions). 
The matrices $c$ and $\Delta d$ include the two-body correlation matrix. 
The matrices in Eq. (\ref{STDDM0}) are given in Appendix B. Equations (\ref{STDDM0}) with $\Delta d = 0$ have been called in the past STDDM (small TDDM) equations \cite{GT}. With inclusion of the nonlinear terms $\Delta d$, we want to call those STDDM* equations.

\subsection{Derivation of STDDM and STDDM* from an Extended Second RPA (ESRPA) and connection with SCRPA}

Let us consider the Equation of Motion (EOM) approach ~\cite{RS,Rowe} with one body and 2-body sectors included fully, that is without restriction on indices (greek labels). We define 
a generalised RPA operator

\begin{equation}
{\mathcal Q}^+_{\nu} = \sum [\chi^{\nu}_{\lambda \lambda'} a^+_{\lambda}a_{\lambda'} + {\mathcal X}^{\nu}_{\lambda_1\lambda_2\lambda'_1\lambda'_2}:a^+_{\lambda_1}
a^+_{\lambda_2}a_{\lambda'_2}a_{\lambda'_1}:]
\label{Q3}
\end{equation}
where $:a^+_{\lambda_1}
a^+_{\lambda_2}a_{\lambda'_2}a_{\lambda'_1}: =a^+_{\lambda_1}
a^+_{\lambda_2}a_{\lambda'_2}a_{\lambda'_1} - [\rho_{\lambda_1 \lambda'_1}\rho_{\lambda_2 \lambda'_2} -  \rho_{\lambda_1 \lambda'_2}\rho_{\lambda_2 \lambda'_1} ]$.
\noindent
As usual with EOM for such an ansatz, we suppose

\[ {\mathcal Q}^+_{\nu}|0\rangle = |\nu\rangle ~~~~~\mbox{and} ~~~~
{\mathcal Q}_{\nu}|0\rangle = 0. \]

\noindent
Minimising the corresponding  energy weighted sum rule $2\Omega_{\nu} = \langle 0|[{\mathcal Q_{\nu}},[H,{\mathcal Q_{\nu}}^+]]|0\rangle /\langle 0|[{\mathcal Q},{\mathcal Q}^+]|0\rangle$, see Sect. V, 
we obtain the following eigenvalue problem

\begin{equation}
\begin{pmatrix}{\mathcal S}&{\mathcal B}\\{\mathcal C}&{\mathcal D}
\end{pmatrix}
\begin{pmatrix} \chi \\ {\mathcal X} \end{pmatrix}
= \Omega
\begin{pmatrix}{\mathcal N}_1&{\mathcal T}\\{\mathcal T}^+&{\mathcal N}_2 \end{pmatrix}
\begin{pmatrix}\chi \\ {\mathcal X} \end{pmatrix},
\label{SCS}
\end{equation}

\noindent
where the various matrix elements are given in an obvious way by the corresponding double commutators (lhs) and commutators (rhs) which correspond to the ones contained in the sum-rule for $\Omega_{\nu}$. The matrix ${\mathcal S}$ contains the correlated occupation numbers and the two-body correlation matrix $C_2$. 
The matrices ${\mathcal B}$, ${\mathcal C}$ and ${\mathcal N}_2$ include the two-body correlation matrix $C_2$ and the three-body correlation matrix $C_3$, and ${\mathcal D}$ can have up to the
four-body correlation matrix $C_4$ though it is neglected.
The matrices in Eq. (\ref{SCS}) are given in \cite{TS10} where this equation was coined ERPA (Extended RPA). 
However, a more appropriate name is 'Extended Second RPA' (ESRPA) because it includes the two body sector and 
reduces to the standard second RPA in the limit where the expectation values are evaluated with the HF state. 
It has been shown in the past that, under certain approximations, this ESRPA is equivalent to the STDDM equation \cite{TS04}.  
Let us sketch this again. For this, in ESRPA, we neglect everywhere $C_3$ (and $C_4$).
This concerns ${\mathcal B}, {\mathcal C}, {\mathcal D}$, and ${\mathcal N}_2$. In ${\mathcal D}$ we additionally neglect the terms which are named in \cite{TTS04} the ${\mathcal T_{32}}$ terms. Those
${\mathcal T_{32}}$ terms correspond to the expectation values of the commutator between two-body and three-body operators \cite{TTS04}.
Then we arrive at the following  structure of above eigenvalue equation (\ref{SCS})

\begin{equation}
\begin{pmatrix}a{\mathcal N}_1+b{\mathcal T}^+&a{\mathcal T}+b{\mathcal N}_2\\c{\mathcal N}_1+ d{\mathcal T}^+&c{\mathcal T}+ d{\mathcal N}_2
\end{pmatrix}
\begin{pmatrix} \chi \\ {\mathcal X} \end{pmatrix}
= \Omega
\begin{pmatrix}{\mathcal N}_1&{\mathcal T}\\{\mathcal T}^+&{\mathcal N}_2 \end{pmatrix}
\begin{pmatrix}\chi \\ {\mathcal X} \end{pmatrix},
\label{SCS'}
\end{equation}

\noindent
where the matrices $a, b, c, d$ are as in (\ref{STDDM0}) (see App. B) 
containing at most $C_2$'s.\\
  Let us notice that in the left matrix the elements [12] and [21] are hermitian conjugates to one another. 
This stems from the fact that already in (\ref{SCS}) the matrices ${\mathcal C}$ and ${\mathcal B}$ are the hermitian conjugates of one another under the condition that they are evaluated 
at equilibrium, see \cite{TS04} for a discussion of this point. The [11] element of the left matrix is also symmetric because at equilibrium we have $i\dot \rho =0$. 
The one body sector of Eq.(\ref{SCS'}) corresponds to Self-Consistent RPA (SCRPA, see below) which was derived independently earlier \cite{NPA628}. 
So, in including correlations, the standard RPA has been upgraded to SCRPA. This is natural because, as mentioned, with correlations the corresponding ground state cannot be the HF state any longer. 
Therefore SCRPA has now found its natural place when the time-dependent HF equations are extended in a consistent way to include two body correlations.  
We will come back to SCRPA in Sect.IV and Sect.V. The [22] element is not hermitian because at this level of our theory we do not fullfill that 3-body and 4-body density matrices are stationary.\\  
Equation (\ref{SCS'}) is intimately related to the STDDM equation as we will show now. 
Defining

\begin{equation}
\begin{pmatrix}\tilde \chi \\ \tilde {\mathcal X} \end{pmatrix}
=
\begin{pmatrix}{\mathcal N}_1&{\mathcal T}\\{\mathcal T}^+&{\mathcal N}_2 \end{pmatrix}
\begin{pmatrix} \chi \\ {\mathcal X} \end{pmatrix}
\label{Trafo}
\end{equation}

\noindent
we obtain the following modified eigenvalue equation

\begin{equation}
\begin{pmatrix}a&b\\c&d
\end{pmatrix}
\begin{pmatrix} \tilde \chi \\ \tilde {\mathcal X} \end{pmatrix}
= \Omega
\begin{pmatrix}\tilde \chi \\ \tilde {\mathcal X} \end{pmatrix}.
\label{STDDM}
\end{equation}

\noindent
The remarkable fact is that this equation is also obtained in linearising around equilibrium the coupled EOM's for $n_{\alpha}$ and $C_2$ as is seen from Eq. (\ref{STDDM0}) without $\Delta d$. 
With the use of Eq. (\ref{Trafo}) the STDDM* equation (Eq. (\ref{STDDM0}) with $\Delta d$ ) can also be expressed as
\begin{equation}
\begin{pmatrix}a{\mathcal N}_1+b{\mathcal T}^+&a{\mathcal T}+b{\mathcal N}_2\\c{\mathcal N}_1+\tilde d{\mathcal T}^+&c{\mathcal T}+\tilde d{\mathcal N}_2
\end{pmatrix}
\begin{pmatrix} \chi \\ {\mathcal X} \end{pmatrix}
= \Omega
\begin{pmatrix}{\mathcal N}_1&{\mathcal T}\\{\mathcal T}^+&{\mathcal N}_2 \end{pmatrix}
\begin{pmatrix}\chi \\ {\mathcal X} \end{pmatrix}.
\label{SCS''}
\end{equation}
Notice that with respect to (\ref{SCS'}) the matrix $d$ is changed into $\tilde d$ in (\ref{SCS''}).
With respect to (\ref{STDDM0}), we want to call the set of equations (\ref{SCS''}), the STDDM*-b equations (or STDDM-b when $\Delta d$ is neglected).
Since ${\mathcal T}$, ${\mathcal N}_2$ and $\Delta d$ contain $C_2$, the [21] and [22] elements of Eq. (\ref{SCS''}) have additional quadratic terms of $C_2$ that correspond to
$C_3$. Thus STDDM*-b is a better approximation to ESRPA than STDDM.  
There is a great consistency between the ESRPA equation and STDDM (and also STDDM*-b). Though STDDM equations of (\ref{STDDM}) and  (\ref{SCS'})  
are equivalent, the explicit form of the two equations (\ref{SCS'}) and (\ref{STDDM}) is quite different in detail.
 For example, it is obvious that neglecting the 2-body amplitudes in STDDM, eq. (\ref{STDDM}), this gives back standard RPA except for partial occupation of the single-particle states. 
 On the other hand, the same reduces STDDM-b ( Eq. (\ref{SCS'})) to SCRPA which contains already, as we will see, much more correlations than standard RPA. 
 In fact, as it was shown in the past \cite{NPA628, NPA512} and will be shown again with the applications below, it can already be a good approximation to STDDM. 
 In a way, it seems natural that in STDDM appears SCRPA. As already discussed, standard RPA corresponds to linearised TDHF. 
 Therefore, the HF Slater determinant is the consistent ground state for standard RPA. Linearised TDDM 
 or STDDM (STDDM*) naturally correspond to a ground state containing correlations. Below, in Sect. V,  we will give as a good approximation  a correlated ground state wave function in terms 
of the one of Coupled Cluster Theory.  \\

It should be noticed that in above STDDM equation $b \ne c^+$ and thus the corresponding matrix is strongly non-symmetric. One, therefore, has to define left and right eigenvectors. 
How this goes in detail is explained in \cite{TS04} where also applications with  good success are presented. On the other hand, (\ref{SCS'}) and (\ref{SCS''}) are  much more symmetric versions of STDDM and STDDM*. The remaining non-hermiticity in the [22] element of the interaction matrix in STDDM* may be eliminated by the prescription of Rowe \cite{Rowe} who explicitly symmetrised the matrix.
If the two versions (\ref{STDDM}) and (\ref{SCS'}) of STDDM (STDDM*, if $\Delta d$ is included) are solved in full, the results will be the same. However, the fact to transform the non-symmetric form of STDDM in (\ref{STDDM}) to the more symmetric STDDM one in (\ref{SCS'}) has apparently transferred a lot of correlations from the 2-body sector to the one body sector (standard RPA vs SCRPA). This may be of importance if in STDDM (or in STDDM*) further approximations are applied. An extreme approximation is to neglect the 2-body amplitudes in both cases where the difference clearly shows up. On the other hand, a non hermitian eigenvalue problem may also entail some problems concerning spurious solutions or non positive definite spectral functions. 
However, in the past applications \cite{Takahara,TS04,Toh07}, this has never caused any serious problems. In a way, the situation is rather similar to the difference which exists between the Dyson boson expansion which leads to a non-hermitian problem and, e.g., the Holstein-Primakoff (Belyaev-Zelevinsky) boson expansion leading to a hermitian matrix \cite{RS}. The basic difference between both methods is, as here, the treatment of the norm matrix.\\

\subsection{Recovering TDDM at equilibrium}

\subsubsection{Equation for $C_{\alpha\beta\alpha'\beta'}$}

WTo show that all equations are consistent, we now want to make a connection of STDDM (and, thus, STDDM*) with TDDM. For this, let us introduce the following identities
supposing $\Omega$ real

\begin{equation}
\sum_{\nu}\bigg [\Omega_{\nu}
\begin{pmatrix}\tilde \chi^{\nu}\\ \tilde {\mathcal X}^{\nu}\end{pmatrix}
(\tilde \chi^{\nu *}~~\tilde {\mathcal X}^{\nu *}) - \begin{pmatrix}\tilde \chi^{\nu}\\ \tilde {\mathcal X}^{\nu}\end{pmatrix}
(\tilde \chi^{\nu *}~~\tilde {\mathcal X}^{\nu *})\Omega_{\nu} \bigg ]
=0.
\label{DM4}
\end{equation}

\noindent
This equation can also be written as

\begin{eqnarray}
 \sum_{\nu}&\bigg [& 
\begin{pmatrix}a&b\\c&d \end{pmatrix}
\begin{pmatrix}\tilde \chi^{\nu}\tilde \chi^{\nu *}&\tilde \chi^{\nu}\tilde {\mathcal X}^{\nu *}\\\tilde {\mathcal X}^{\nu}\tilde \chi^{\nu *}&\tilde {\mathcal X}^{\nu}\tilde {\mathcal X}^{\nu *}\end{pmatrix}\nonumber\\
&-&\begin{pmatrix}\tilde \chi^{\nu}\tilde \chi^{\nu *}&\tilde \chi^{\nu}\tilde {\mathcal X}^{\nu *}\\\tilde {\mathcal X}^{\nu}\tilde \chi^{\nu *}&\tilde {\mathcal X}^{\nu}\tilde {\mathcal X}^{\nu *}\end{pmatrix}\begin{pmatrix}a&c^+\\b^+&d\end{pmatrix}
\bigg  ] =0.
\label{DM5}
\end{eqnarray}

\noindent
The one body sector of this equation is the following (please note that in some earlier publications the definitions of the $b$ and $c$ matrices has been inversed, see, e.g., \cite{TS04}).

\begin{equation}
\sum_{\nu}[a\tilde \chi^{\nu}\tilde \chi^{\nu *}+b\tilde {\mathcal X}^{\nu}\tilde \chi^{\nu *} -\tilde \chi^{\nu}\tilde \chi^{\nu *}a
- \tilde \chi^{\nu}\tilde {\mathcal X}^{\nu *} b^+] =0.
\label{1b-sct}
\end{equation}

\noindent
With the following identity

\begin{eqnarray}
\sum_{\nu}\tilde {\mathcal X}^{\nu}_{\alpha \beta \alpha' \beta'}\tilde{\chi}^{\nu *}_{\gamma'\gamma}&=&\delta_{\alpha\gamma'}
C_{\gamma \beta \alpha' \beta'}-\delta_{\beta\gamma'}C_{\gamma \alpha \alpha' \beta'} 
\nonumber \\
&+& n_{\gamma\alpha'}C_{\alpha\beta\beta'\gamma'} - n_{\gamma\beta'}C_{\alpha\beta\alpha'\gamma'}
\nonumber \\
&-&n_{\beta\gamma'}C_{\alpha\gamma\alpha'\beta'}+n_{\alpha\gamma'}C_{\beta\gamma\alpha'\beta'}
\nonumber \\
&+&C_{\alpha \beta \gamma \alpha' \beta' \gamma'},
\label{3b-C3}
\end{eqnarray}

\noindent
we obtain, specialising to $p$ and $h$ indices the static form of the TDDM equations (\ref{static}) - (\ref{T4h}) which should be  {\it unique} equations for the $C_2$'s. 
It is interesting to note that if one restricts the $C_3$ to $p, h$ indices only and also keeping only $\chi_{ph}$ or $\chi_{hp}$ components, then ${\mathcal X}$ amplitudes can only be of the 3$p$-1$h$ (3$h$-1$p$) type. This is consistent with the extended RPA operator treated in Sect. V where the two body sector also only contains 3$p$-1$h$ (3$h$-1$p$) amplitudes.
\\

\subsubsection{Occupation probability from ESRPA}

 We have shown that Eq. (\ref{1b-sct}) is equivalent to the stationary condition of Eq. (\ref{N3C2}). Now we must consider how the occupation probability $n_{\alpha}$ 
is expressed by the transition amplitudes in ESRPA.
We assume the following relation for the diagonal occupation matrix $\rho_{\alpha\alpha'}=n_\alpha\delta_{\alpha\alpha'}$ 
\begin{eqnarray}
\sum_{\nu}\tilde{\chi}_{\alpha\alpha'}^\nu\tilde{\chi}_{\beta'\beta}^{\nu*}&=&
\sum_{\nu}\langle 0|a^+_{\alpha'}a_\alpha|\nu\rangle\langle\nu|a^+_{\beta'}a_\beta|0\rangle
\nonumber \\
&=&\delta_{\alpha\beta'}\langle 0|a^+_{\alpha'}a_\beta|0\rangle+\langle 0|a^+_{\alpha'}a^+_{\beta'} a_\beta a_\alpha|0\rangle
\nonumber \\
&=&\delta_{\alpha\beta'}\delta_{\beta\alpha'}n_\beta\bar{n}_\alpha +C_{\alpha\beta\alpha'\beta'},
\label{D0}
\end{eqnarray}
where $\bar{n}_\alpha=1-n_\alpha$.
From Eq. (\ref{D0}) we obtain
\begin{eqnarray}
\sum_{\nu}\tilde \chi^\nu_{\alpha\alpha}\tilde \chi^{\nu *}_{\alpha\alpha}=n_{\alpha}(1-n_{\alpha})+C_{\alpha\alpha\alpha\alpha}
&=&n_{\alpha}-n_{\alpha}^2.
\end{eqnarray}
The above equation gives for the occupation numbers
\begin{eqnarray}
n_{\alpha}=\frac{1}{2}\left(1\pm\sqrt{1-4\sum_{\nu\neq 0}\tilde \chi^\nu_{\alpha\alpha}\tilde \chi^{\nu *}_{\alpha\alpha}}\right).
\label{ninSTDDM}
\end{eqnarray}
In RPA and SCRPA there is no diagonal one-body amplitude such as $ \chi^\nu_{\alpha\alpha}$, whereas in ESRPA $\chi^\nu_{\alpha\alpha}$ can couple to ${\mathcal X}^\nu_{\alpha\beta\alpha'\beta'}$ 
which has the same quantum numbers as the ground state. Thus the occupation probabilities in ESRPA are determined by two-phonon states expressed by ${\mathcal X}^\nu_{\alpha\beta\alpha'\beta'}$,
which is in contrast with SCRPA.   
We use Eq. (\ref{ninSTDDM}) to calculate the occupation probabilities in ESRPA. 
Let us notice that relation (\ref{ninSTDDM}) has the same structure as the occupation numbers obtained from BCS theory when expressed via the BCS amplitudes $v_iu_i = \kappa_i$ \cite{RS}. 

\subsubsection{Correlation energy from $C_{\alpha\beta\alpha'\beta'}$} 

Usually, the correlation energy is defined as the difference of the total correlated energy minus the Hartree-Fock energy. In this work, we thought it more appropriate to consider what one could call the 2-body correlation energy (for example in the case of BCS theory, this would reduce to the pairing energy) $E_{\rm 2bcor}$ 
defined by
\begin{eqnarray}
E_{\rm 2bcor}=\frac{1}{4}\sum_{\alpha\beta\alpha'\beta'}\bar{v}_{\alpha\beta\alpha'\beta'}C_{\alpha'\beta'\alpha\beta}.
\label{ecorr}
\end{eqnarray}
The equation for $\tilde \chi^\nu_{\alpha\alpha'}$ in STDDM, $a\tilde \chi^\nu+b\tilde {\mathcal X}^\nu=\Omega_\nu \tilde \chi^\nu$, gives
\begin{eqnarray}
\Omega_\nu \tilde \chi^\nu_{\alpha\alpha'}&=&(\epsilon_{\alpha}-\epsilon_{\alpha'})\tilde \chi^\nu_{\alpha\alpha'}
\nonumber \\
&+&(n_{\alpha'}-n_{\alpha})\sum_{\lambda\lambda'}\bar{v}_{\alpha\lambda'\alpha'\lambda}\tilde \chi^\nu_{\lambda\lambda'}
\nonumber \\
&+&\frac{1}{2}\sum_{\lambda_1\lambda_2\lambda_3}(
\bar{v}_{\alpha\lambda_1\lambda_2\lambda_3}\tilde {\mathcal X}^\nu_{\lambda_2\lambda_3\alpha'\lambda_1}
\nonumber \\
&-&\bar{v}_{\lambda_1\lambda_2\alpha'\lambda_3}\tilde {\mathcal X}^\nu_{\alpha\lambda_3\lambda_1\lambda_2})
\label{Eint0}
\end{eqnarray}
Multiplying $\tilde \chi^{\nu *}_{\beta'\beta}$ and using Eqs. (\ref{3b-C3}) and (\ref{D0}), we obtain
\begin{eqnarray}
\sum_{\mu \alpha}\Omega_\nu \tilde \chi^\nu_{\alpha\alpha}\tilde \chi^{\nu *}_{\alpha\alpha}&=&\sum_{\alpha\lambda\lambda'}\bar{v}_{\alpha\lambda\alpha\lambda'}C_{\alpha\lambda'\alpha\lambda}
\nonumber \\
&-&\frac{1}{2}\sum_{\alpha\lambda\lambda'\lambda''}\bar{v}_{\lambda\lambda'\alpha\lambda''}C_{\alpha\lambda''\lambda\lambda'}.
\label{Eint4}
\end{eqnarray}
The first term on the right-hand side has no contribution in the solvable models discussed below. 
In general, $C_{phph'}$, $C_{pp'pp''}$, $C_{hphp'}$ and
$C_{hh'hh''}$ are smaller than $C_{pp'hh'}$ and $C_{hh'pp'}$ in a perturbative regime. Therefore, $E_{\rm 2bcor}$ can approximately be expressed as 
\begin{eqnarray}
E_{\rm 2bcor}\approx-\frac{1}{2}\sum_{\nu \alpha}\Omega_\nu \tilde \chi^\nu_{\alpha\alpha}\tilde \chi^{\nu *}_{\alpha\alpha}.
\label{Eint3}
\end{eqnarray}
Equation (\ref{Eint3}) has only diagonal elements $\tilde \chi^\nu_{\alpha\alpha}$, what means that in ESRPA $E_{\rm 2bcor}$ is determined by two-phonon states similarly to
the occupation probabilities (Eq. (\ref{ninSTDDM})).
We calculate $E_{\rm{2bcor}}$ in ESRPA using Eq. (\ref{Eint3}). It will also be the expression we use for the applications in Sect.VII. Since with (\ref{ninSTDDM}) we have the occupation numbers, we can also calculate the one body part of the energy and, thus, the total energy is given as well.

\noindent
\section{Self-Consistent RPA in relation with TDDM}

\subsection{General case}

As we have mentioned, the one body sector of STDDM-b and STDDM*-b is equivalent to what is known in the literature as SCRPA. Because the one body sector is of importance for applications but also in its own right, we, for completeness, will again dwell on it in this and the next section. However, the reader already familiar with SCRPA, or not specially interested in this issue, can directly jump to the applications, section VI.\\ 

Let us start writing down the most general single particle RPA operator as

\begin{equation}
Q^+_{\nu} = \sum_{\alpha \beta, \alpha \ne \beta}\chi^{\nu}_{\alpha \beta}a^+_{\alpha}a_{\beta},
\label{RPAop}
\end{equation}

\noindent
where, as usual, 

\begin{equation}
|\nu \rangle = Q^+_{\nu}|0\rangle
\label{nu}
\end{equation}

\noindent
is the excited state. The RPA operator also is supposed to possess the killing property (see Sect. V)

\begin{equation}
Q_{\nu}|0\rangle = 0.
\label{kill}
\end{equation}

\noindent
We can define an average excitation energy using the energy weighted sum rule

\begin{equation}
\Omega_{\nu} = \frac{1}{2}\frac{\langle 0|[Q_{\nu},[H,Q^+_{\nu}]]|0\rangle}{\langle 0|[Q_{\nu},Q^+_{\nu}]|0\rangle }.
\label{sumrule}
\end{equation}

\noindent
Varying $\Omega_{\nu}$ with respect to the amplitudes $\chi_{\alpha \beta}$ leads to the following eigen value problem

\begin{eqnarray}
{\mathcal S}\chi^{\mu} =\Omega_{\mu}{\mathcal N}_{1}\chi^{\mu},
\label{scrpa} 
\end{eqnarray}
where 
\begin{eqnarray}
&{\mathcal S}&(\alpha\alpha':\lambda\lambda')= \langle 0|[a^+_{\alpha'}a_{\alpha},[H,a^+_{\lambda}a_{\lambda'}]]|0\rangle \nonumber\\
&=&
(\epsilon_\alpha-\epsilon_{\alpha'})
(n_{\alpha'}
-n_{\alpha})\delta_{\alpha\lambda}\delta_{\alpha'\lambda'}
\nonumber \\
&+&(n_{\alpha'}-n_{\alpha})(n_{\lambda'}-n_{\lambda})\bar{v}_{\alpha\lambda'\alpha'\lambda}
\nonumber \\
&-&\delta_{\alpha'\lambda'}\frac{1}{2}\sum_{\gamma\gamma'\gamma''}\bar v_{\alpha\gamma \gamma'\gamma''}
C_{\gamma'\gamma''\lambda\gamma}
\nonumber \\
&-&\delta_{\alpha\lambda}\frac{1}{2}\sum_{\gamma\gamma'\gamma''}\bar v_{\gamma\gamma'\alpha'\gamma''}
C_{\lambda'\gamma''\gamma\gamma'}
\nonumber \\
&+&\sum_{\gamma\gamma'}(\bar v_{\alpha\gamma \lambda\gamma'}
C_{\lambda'\gamma'\alpha'\gamma}
+\bar v_{\lambda'\gamma \alpha'\gamma'}
C_{\alpha\gamma'\lambda\gamma})
\nonumber \\
&-&\frac{1}{2}\sum_{\gamma\gamma'}(\bar v_{\alpha\lambda'\gamma\gamma'}
C_{\gamma\gamma'\alpha'\lambda}
+\bar v_{\gamma\gamma'\alpha'\lambda}
C_{\alpha\lambda'\gamma\gamma'}),
\label{S-term}
\\
&{\mathcal N}&_{1}(\alpha\alpha':\lambda\lambda')=(n_{\alpha'}
-n_{\alpha})\delta_{\alpha\lambda}\delta_{\alpha' \lambda'}.
\label{snorm}
\end{eqnarray}
If we replace the RPA ground state by the HF one, then the matrix ${\mathcal S}$ reduces to the HF stability matrix  and ${\mathcal N}^{(0)}_1$ becomes the metric matrix of RPA \cite{RS} and, thus, the standard RPA equations are recovered. 
The normalisation of the amplitudes $\chi^{\nu}_{\alpha \beta}$ is given by

\begin{equation}
\sum \chi^{\nu *}_{\alpha \beta}{\mathcal N}_{1}(\alpha\alpha':\lambda\lambda')\chi^{\nu'}_{\lambda \lambda'}=\delta_{\nu,\nu'},
\label{normalisation}
\end{equation}
where $\chi^{\nu *}_{\alpha \beta}$ is the left eigenvector.
Above eigenvalue problem is equivalent to SCRPA \cite{NPA628} with amplitudes $\chi_{\alpha \beta}$ where there are no restrictions on the indices 
besides $\alpha \ne \beta$. This stems from the fact that ${\mathcal N}_1$ acts as a norm matrix like it appears in problems where one works with a non-orthonormal basis \cite{RS}. In such cases, in general, one has to diagonalise the norm matrix  and divide the Hamilton matrix from left and right with the  the square roots of the eigenvalues. Configurations with zero (or near zero) eigenvalues have to be excluded for obvious reasons. In the SCRPA case, this just happens for diagonal, or nearly diagonal amplitudes $ \chi_{\alpha \alpha}$ which, thus, cannot be included. This can only be done, as we discussed before, if the two particle sector is also considered. \\

The fact that the diagonal amplitudes cannot be included in (\ref{scrpa}), allows us to rewrite this equation in a form which has the mathematical structure of standard RPA. To this end, we re-write the RPA excitation operator (\ref{RPAop}) in a somewhat different form

\begin{equation}
Q^+_{\nu} = \sum_{k_1>k_2}(X^{\nu}_{k_1k_2}\delta Q^+_{k_1k_2} - Y^{\nu}_{k_1k_2}\delta Q_{k_1k_2})
\label{RPAop-new}
\end{equation}
with

\begin{equation}
\delta Q^+_{k_1k_2} = N^{-1/2}_{k_1k_2}a^+_{k_1}a_{k_2}
\label{dQ}
\end{equation}
and

\begin{equation}
N^{1/2}_{k_1k_2} = \sqrt{n_{k_2} - n_{k_1}}.
\label{k1>k2}
\end{equation}

This leads straightforwardly to the following RPA eigenvalue problem

\begin{equation}
\begin{pmatrix} A&B\\-B^*&-A^*
\end{pmatrix}
\begin{pmatrix}X\\Y\end{pmatrix}
=\Omega_{\nu}\begin{pmatrix}X\\Y\end{pmatrix}
\label{SCRPA}
\end{equation}

\noindent
with

\begin{eqnarray}
A_{k_1k_2,k'_1k'_2} &=& \langle [ \delta Q_{k_1k_2},[H,\delta Q^+_{k'_1k'_2}]]\rangle \nonumber \\
B_{k_1k_2,k'_1k'_2} &=& -\langle [\delta Q_{k_1k_2},[H,\delta Q_{k'_1k'_2}]]\rangle. 
\label{AB}
\end{eqnarray}

The $X, Y$ amplitudes have the usual orthonormalisation relations of standard $ph$-RPA with the replacements $p \leftrightarrow k_1$ and $h \leftrightarrow k_2$. Of course the $A$ and $B$ matrices are closely related to the ${\mathcal S}$ matrix of (\ref{S-term}).

In order to calculate the $C_2$ correlation functions entering the SCRPA matrix, one can either get them from the static solution of the TDDM equations with quadratic decoupling of $C_3$ with respect to the $C_2$'s (this will later be called the C-RPA scheme) or one  establishes a selfconsistent cycle, for which we must give a relation between $C_2$ and the RPA amplitudes $X, Y$. For this, it is convenient to
introduce the 'bosonic' density matrix ${\mathcal R}$ 

\begin{equation}
{\mathcal R} =  \sum_{\nu}\begin{pmatrix}Y^{\nu*}Y^{\nu}&Y^{\nu*}X^{\nu}\\X^{\nu*}Y^{\nu}&X^{\nu*}X^{\nu}\end{pmatrix}\equiv \begin{pmatrix}R&K\\K^+&1+R^+ \end{pmatrix},
\label{DM}
\end{equation}
with $({\mathcal N}_0{\mathcal R})^2 = - {\mathcal N}_0{\mathcal R}$ where ${\mathcal N}_0$ $=$ $\begin{pmatrix}1&0\\0&-1\end{pmatrix}$

\noindent
and where we can make the following identifications 

\begin{eqnarray}
&R&_{k_1k_2k'_1k'_2} \equiv N^{-1/2}_{k_1k_2}[n_{k_2}\bar n_{k_1} + C_{k_1k'_2k_2k'_1}]N^{-1/2}_{k'_1k'_2} \nonumber\\
&K&_{k_1k_2k'_1k'_2} \equiv N^{-1/2}_{k_1k_2}C_{k_1k'_1k_2k'_2}N^{-1/2}_{k'_1k'_2}.
\label{R,K}
\end{eqnarray}

It may be interesting to rewrite the RPA equations in still a different form. With 

\begin{equation}
\begin{pmatrix} A&B\\-B^*&-A^*
\end{pmatrix}
\equiv  {\mathcal H}
\label{SCRPA-matrix}
\end{equation}
we can write (\ref{SCRPA}) as 

\begin{equation}
{\mathcal R} {\mathcal H}^+ -  {\mathcal H}{\mathcal R} = 0
\label{R,H}
\end{equation} 

This form reminds the BCS (or HFB) equations of superconductivity \cite{RS} with, however some different signs due to the bosonic structure of the RPA equations.
The introduction of the density matrix ${\mathcal R}$ has the advantage that one easily can restore a missing antisymmetry as we will see in Section IV.C.\\

It remains to express the occupation numbers in terms of the RPA amplitudes to establish  fully self-consistent RPA equations. Because of the Fermi surface the occupation numbers can be divided in hole and particle occupancies $n_h$ and $n_p$. How the latter are connected to $C_2$'s and, thus, to the RPA amplitudes will be shown in Sect. V.

\begin{figure}
\includegraphics[width=8.6cm]{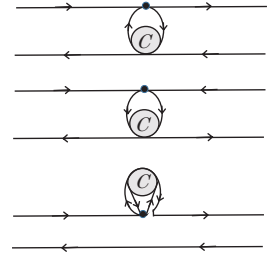}
\caption{\label{scrpa-graph1}
Screening terms and self-energy corrections. Symmetric graphs exist where the interaction (full dot) is attached to the hole line (arrow to the left)}
\end{figure}

At this point, it may be appropriate to interpret the different terms of the $A$ and $B$-matrices. The standard terms are, of course, trivial and have been discussed in text books \cite{RS}. The other terms are displayed graphically in  Fig.\ref{scrpa-graph1}. Analogous graphs exist (not displayed) where the interaction (full dot) is attached to the hole line with arrow to the left. 
Their interpretation is clear. The first two terms constitute instantaneous $ph$ and $pp(hh)$ exchange terms with respect to the external $p$ and $h$ lines. They, therefore, screen (eventually anti screen) the bare interaction. Such screening terms have been discussed in the literature since very long. The iteration of the equations gives raise to so-called 'bubble into bubble' terms \cite{Kirson}. The particularity of our formalism here is that those terms emerge from a general formalism and that they are {\it instantaneous}. They can, therefore, be incorporated into standard RPA programs. The third term in Fig.\ref{scrpa-graph1} obviously corresponds to a self-energy correction due to RPA modes. 
Those correspond to the famous particle vibration corrections to the mean field. Again the particularity here is that this correction is instantaneous.\\

For the solution of the SCRPA equations, several routes are possible. The standard way is to express the correlation functions with the $X$ and $Y$ amplitudes as discussed just above.  With the present formalism one also can evaluate the correlation functions $C_2$'s either from (\ref{R,H}) or below from (\ref{gs}) and then insert them into the $A$ and $B$ matrices. Also the single particle occupancies can be included in this way via eq (\ref{nh}), see Sect. V. The results will depend on  whether we take the non antisymmetrised or the antisymmetrised form of ${\mathcal R}$. Only the non-antisymmetrised form will be equivalent to the standard way in expressing everything by the $X, Y$ amplitudes. We will come back to this with the applications. A further possibility is to take the $C_2$'s directly from the static limit of the TDDM equations quadratic in the $C_2$'s. As mentioned, we call this the C-RPA (correlated RPA). We will see with the applications in Sect.VII that all these variants give quite close answers at least up to coupling strengths where the standard HF equations become unstable indicating that the system undergoes a phase transition.\\

\subsection{ Properties of SCRPA}

Before we go into the details of how SCRPA is connected with  TDDM, let us outline some properties of SCRPA. One of the most important ones, fulfilled by the standard RPA, is the so-called energy weighted sum-rule.\\

\begin{equation}
S_1 = \sum_{\nu}\Omega_{\nu}|\langle \nu|F|0 \rangle|^2 = \frac{1}{2}\langle 0|[F,[H,F]]|0\rangle
\end{equation}

\noindent
where $F=\sum_{\alpha \beta}f_{\alpha \beta} a^+_{\alpha}a_{\beta}$ is supposed to be a hermitian one body (excitation) operator. Then for the right hand side we can write

\begin{eqnarray}
S_1 &=& \frac{1}{2}\sum_{\nu} Tr[f^+\chi^{\nu}S\chi^{\nu,+}f] \nonumber\\
&=& \frac{1}{2}Tr[f^+\sum_{\nu}\Omega_{\nu}{\mathcal N}_1\chi^{\nu,+}f] 
\nonumber \\
&=& \sum_{\nu}\Omega_{\nu}|\langle 0|F|\nu \rangle|^2
\end{eqnarray}

\noindent
Therefore also SCRPA fulfills the f-sum rule. This has, e.g., been discussed in  \cite{Cat96}, \cite{Colloquium}. From the fulfillment of the sum rule, it also follows that the Goldstone theorem is satisfied. For example in nuclear physics the translational motion is always broken, if one works in a localised single particle basis. Then the SCRPA separates the so-called spurious mode  at zero energy, if the single particle basis is chosen from the generalised mean field equation $\langle 0|[H,Q^+_{\nu}]|0 \rangle =\langle 0|[H,a^+_{\alpha'}a_{\alpha}]|0 \rangle =0 $ which is the static limit of (\ref{n}) ~\cite{Colloquium}. The fullfillment of sum rule and Goldstone mode stems from the fact that the RPA operator (\ref{RPAop}) contains all types of indices, that is not only $ph$ but also $pp$ and $hh$ ones. The RPA operator (\ref{RPAop}) contains as a particular case, e.g., the total momentum operator $\hat P$ which commutes with the Hamiltonian. From (\ref{sumrule}) we then see that the zero mode appears. 
The fullfillment of the Goldstone theorem has already explicitly been demonstrated in \cite{epja32,Lipkin3}. Consequently SCRPA as defined in this section has some important properties in common with standard RPA.  This is a very rewarding feature because generally it is not easy to set up a practical scheme, going beyond standard RPA, which obeys conservation laws, sum rules, and Goldstone theorem. However SCRPA is an approximation to STDDM-b (or STDDM*-b) and, therefore, also fails in some respects. For example in the superfluid (superconducting) case, the symmetry operator is the particle number operator which in the quasiparticle basis has a diagonal (hermitian) piece. This cannot be included into SCRPA because the norm matrix (\ref{snorm}) has a zero eigenvalue. Thus, self-consistent quasi-particle RPA will not give the zero or Goldstone mode. For this the consideration of the STDDM approach is necessary. It may, however, be possible to include the 2-body sector only in approximate form, that is eventually to lowest order perturbation theory.\\ 
Another important property of standard RPA which is fulfilled by SCRPA is gauge invariance. Gauge invariance of standard RPA is nicely demonstrated by Feldman and Fulton \cite{FF84}. The extra terms containing the two body correlation functions in (\ref{S-term}) cancel in the limit where the two open legs are put on the same spot in position space. Actually, gauge invarince of standard RPA {\it as well as} SCRPA can easily be verified from (\ref{AB}). If in these equations the operator $\delta Q_{\alpha \beta}$ is transformed into $r$-space and the diagonal element is taken, as demanded to show gauge invariance (see \cite{FF84}, Eq. (3.69)), we immediately realise that this diagonal operator commutes with the remainder (also written in $r$ space), once the Hamiltonian $H$ is replaced by its interaction part $V$, that is, the Coulomb interaction. Therefore,  gauge invariance is fullfilled. This argument is valid discarding spin but, as shown in \cite{FF84}, this does not invalidate the general proof. 
These considerations also entail that the so-called 'velocity-length' equivalence in the dipole transition is preserved \cite{FF84}, see also \cite{Colloquium}.

\subsection{SCRPA-content of TDDM}

Let us now investigate  how much of TDDM is already incorporated in SCRPA. To this end,
we may consider the equation for the antisymmetrised density matrix $\tilde {\mathcal R}_{\alpha\beta\alpha'\beta'}$, rather than the non-antisymmetrized one of (\ref{DM}) (see (\ref{Trafo}) for the definition of $\tilde \chi$)
\begin{eqnarray}
&\tilde {\mathcal R}&_{\alpha\beta\alpha'\beta'}=
\frac{1}{2}\sum_{\nu}(\tilde{\chi}_{\alpha\alpha'}^{\nu}\tilde{\chi}_{\beta'\beta}^{\nu*}-\tilde{\chi}_{\beta\alpha'}^{\nu}\tilde{\chi}_{\beta'\alpha}^{\nu*})
\nonumber \\
&=&\frac{1}{2}(\delta_{\alpha\beta'}\delta_{\beta\alpha'}n_\beta\bar{n}_\alpha-\delta_{\alpha\alpha'}\delta_{\beta\beta'}n_\alpha\bar{n}_\beta)
+C_{\alpha\beta\alpha'\beta'}
\label{D}
\end{eqnarray}
to derive an equation for $C_{\alpha\beta\alpha'\beta'}$.

Using
\begin{eqnarray}
&\sum_\nu&(\Omega_\nu {\mathcal N}_1\chi^{\nu}\tilde{\chi}^{\nu*}-\tilde{\chi}^{\nu}{\chi}^{\nu*}{\mathcal N}_1\Omega_\nu)\nonumber\\
&=&\sum_\mu({\mathcal S}\chi^{\nu}\tilde{\chi}^{\nu*}-\tilde{\chi}^{\nu}{\chi}^{\nu*}{\mathcal S}) \nonumber\\
&=&\sum_\mu({\mathcal S}{\mathcal N}^{-1}_1{\mathcal N}_1\chi^{\nu}\tilde{\chi}^{\nu*}-\tilde{\chi}^{\nu}{\chi}^{\nu*}{\mathcal N}_1{\mathcal N}_1^{-1}{\mathcal S})
\nonumber \\
&=&\sum_\mu({\mathcal S}{\mathcal N}^{-1}_1\tilde{\chi}^{\nu}\tilde{\chi}^{\nu*}-\tilde{\chi}^{\nu}\tilde{\chi}^{\nu*}{\mathcal N}_1^{-1}{\mathcal S}), 
\label{D1}
\end{eqnarray}
we can express the equation for $\tilde {\mathcal R}$ as $\tilde {\cal H}\tilde {\mathcal R}-\tilde {\mathcal R} \tilde {\cal H}^+=0$,
where $\tilde {\cal H}={\mathcal S}{\mathcal N}^{-1}_1$.

The explicit expression for $\tilde{\cal H}\tilde {\mathcal R}- \tilde {\mathcal R} \tilde{\cal H}^+=0$ is
\begin{eqnarray} 
&(&\epsilon_\alpha+\epsilon_\beta-\epsilon_{\alpha'}-\epsilon_{\beta'})C_{\alpha\beta\alpha'\beta'}\nonumber\\
&+&\bar v_{\alpha\beta\alpha'\beta'}
(\bar{n}_\alpha \bar{n}_\beta {n}_{\alpha'}{n}_{\beta'}-n_\alpha n_\beta \bar{n}_{\alpha'}\bar{n}_{\beta'})
\nonumber \\
&+&\frac{1}{2}\sum_{\lambda\lambda'}[\bar v_{\alpha\lambda'\alpha'\lambda}(n_{\alpha'}-n_\alpha)C_{\lambda\beta\lambda'\beta'}\nonumber\\
&+&\bar v_{\beta\lambda'\alpha'\lambda}(n_{\alpha'}-n_\beta)C_{\alpha\lambda\lambda'\beta'}
\nonumber \\
&-&\bar v_{\beta\lambda'\beta'\lambda}(n_{\beta'}-n_{\beta})C_{\alpha\lambda\lambda'\alpha'}\nonumber\\
&+&\bar v_{\alpha\lambda'\beta'\lambda}(n_{\beta'}-n_{\alpha})C_{\beta\lambda\lambda'\alpha'}]
\nonumber \\
&+&E_{\alpha\beta\alpha'\beta'}+F_{\alpha\beta\alpha'\beta'}+G_{\alpha\beta\alpha'\beta'}=0.
\label{gs}
\end{eqnarray}
We see that (\ref{gs}) has a similar structure as (\ref{static}). In (\ref{gs}),
the second term corresponds to $B^0_{\alpha\beta\alpha'\beta'}$ and the third term to $H^0_{\alpha\beta\alpha'\beta'}$ except for a factor $1/2$. The additional factor 1/2 is contained in the $F$ matrix, see App. C where
the matrices $E_{\alpha\beta\alpha'\beta'}$, $F_{\alpha\beta\alpha'\beta'}$ and $G_{\alpha\beta\alpha'\beta'}$ are given.\\

\vspace{1cm}

Since the expressions for $E, F, G$ are rather lengthy due to a somewhat complicated structure of how the single particle occupation factors enter, we want to simplify the analysis and replace the occupation numbers by their free, i.e., mean field values $n^0_{\alpha}$. This will be sufficient to show that SCRPA also contains quadratic forms in $C_2$'s quite analogous to TDDM. Taking the free occupation numbers automatically projects all quantities to have $p$ or $h$ indices only.
It can be verified that in this way from (\ref{gs}), the TDDM Eqs. (\ref{static}) are fully recovered up to the linear terms in $C_2$'s. Some differences appear in the quadratic expressions. They are contained in the $G$ matrix in (\ref{gs}). Therefore, let us make some comparisons between the {\it static} TDDM and the above defined form of SCRPA.

\subsubsection{Special cases}

\noindent
i) {\bf 2p-2h configurations}: $T_{p_1p_2h_1h_2}$\\

Here we consider Eq.(\ref{gs}) for $C_{p_1p_2h_1h_2}$ assuming $n_\alpha=1$ or $0$. 
Equation (\ref{E}) has no contribution for $C_{p_1p_2h_1h_2}$ and it is easy to check that the terms in Eq.(\ref{gs}) except for $G_{\alpha\beta\alpha'\beta'}$
are the same as $B^0_{\alpha\beta\alpha'\beta'}$, $P^0_{\alpha\beta\alpha'\beta'}$ and $H^0_{\alpha\beta\alpha'\beta'}$ for $C_{p_1p_2h_1h_2}$. Therefore,
we investigate only the terms with the three-body correlation matrix in (\ref{static}).
The TDDM equation gives Eq. (\ref{T2p2h}).
Only the terms in the first two sums in Eq. (\ref{G}) contribute to $G_{p_1p_2h_1h_2}$ and it is written as
\begin{eqnarray}
G_{p_1p_2h_1h_2}&=&-\frac{1}{2}\sum_{pp'hh'}[\bar v_{ p_1phh'} C_{hh'p'p}
C_{p'p_2h_1h_2}
\nonumber \\
&-&\bar v_{ p_2phh'} C_{hh'p'p}
C_{p'p_1h_1h_2}
\nonumber \\
&+&\bar v_{ pp'h_1h} C_{h'hpp'}
C_{p_1p_2h'h_2}
\nonumber \\
&-&\bar v_{ pp'h_2h} C_{h'hpp'}C_{p_1p_2h_1h'}].
\label{G1}
\end{eqnarray}
Therefore, Eqs. (\ref{T2p2h}) and (\ref{G1}) agree with each other.\\

\noindent
ii) {\bf ph-ph configurations}: $T_{p_1h_1p_2h_2}$\\

The TDDM equation gives Eq. (\ref{Tph}).
For $G_{p_1h_1p_2h_2}$ we obtain
\begin{eqnarray}
G_{p_1h_1p_2h_2}&=&\frac{1}{2}\sum_{pp'hh'}\bar v_{ p_1hh'p} C_{p'ph_2h}
C_{h_1h'p_2p'}
\nonumber \\
&+&\frac{1}{4}\sum_{pp'p''h}\bar v_{ p_1pp'p''} C_{p'p''hh_2}
C_{h_1hp_2p}
\nonumber \\
&-&\frac{1}{2}\sum_{pp'hh'}\bar v_{ ph'p_2h} C_{hh_1pp'}
C_{p_1p'h'h_2}
\nonumber \\
&-&\frac{1}{4}\sum_{pp'p''h}\bar v_{ p'p''p_2p} C_{p_1ph_2h}
C_{hh_1p'p''}
\nonumber \\
&-&\frac{1}{2}\sum_{pp'hh'}\bar v_{ h_1pp'h} C_{h'hp_2p}
C_{p'p_1h'h_2}
\nonumber \\
&+&\frac{1}{4}\sum_{phh'h''}\bar v_{h_1hh'h''} C_{h'h''p_2p}
C_{pp_1hh_2}
\nonumber \\
&+&\frac{1}{2}\sum_{pp'hh'}\bar v_{ph'h_2p'} C_{p_1p'hh'}
C_{h_1hp_2p}
\nonumber \\
&-&\frac{1}{4}\sum_{phh'h''}\bar v_{h'h''hh_2} C_{pp_1h'h''}
C_{h_1hp_2p}.
\label{G2}
\end{eqnarray}
There is a factor of 2 difference between Eqs.(\ref{Tph}) and (\ref{G2}).
If $C_{php'h'}$ is also included, $G_{p_1h_1p_2h_2}$ has still more terms.\\

\noindent
iii) {\bf 4p and 4h configurations}: $ T_{p_1p_2p_3p_4}, T_{h_1h_2h_3h_4} $\\

The TDDM equation for $C_{p_1p_2p_3p_4}$ is given by Eq. (\ref{T4p})
In the case of Eq. (\ref{gs}) $G_{p_1p_2p_3p_4}=0$ and $G_{h_1h_2h_3h_4}=0$ and, thus, this leads to another difference with the TDDM equations.

\subsubsection{Summary of special cases}

1) The equation for $C_{p_1p_2h_1h_2}$ derived from SCRPA has one to one correspondence with TDDM
except for the coupling to $C_{p_1p_2p_3p_4}$ and $C_{h_1h_2h_3h_4}$.
However, there is a factor of 1/2 difference in the $C_2^2$ terms when $C_{p_1h_1p_2h_2}$ is included.

2) The equation for $C_{p_1h_1p_2h_2}$ derived from SCRPA has always a factor of 1/2 difference with TDDM 
if we assume the symmetry $C_{p_1h_1p_2h_2}=-C_{h_1p_1p_2h_2}$. It was, however,  discussed in \cite{TS14} that most of the time $C_{php'h'}$ is smaller than $C_{pp'hh'}$ and then this difference between TDDM and SCRPA will not show up strongly.\\

Therefore at equilibrium, we get with SCRPA very similar equations for the $C_2$'s as with static TDDM. Notably the terms quadratic in $C_2$'s are quite analogous in both cases. There are some differences, however. First comes the fact that, as mentioned, the terms $C_{php'h'}$ are missing factors of two. They are, however, usually smaller than the $C_{pp'hh'}$ and, then, this should not affect the results very much. There is, however, another difference between TDDM and SCRPA. This concerns the fact that on the r.h.s. of (38) there is the norm matrix which leads to the division by $n_{\beta} - n_{\alpha}$ in $\tilde{\mathcal H} = {\mathcal S}{\mathcal N}^{-1}_1$ of (43). The significance of this for the ground state is not very evident and the occupation factors can be replaced by the free ones to good approximation. However, for excited states it may be very important to keep the correlated $n_{\alpha}$'s, since the norm matrix serves to make out of the non-orthonormalised basis in (34) an orthonormalised one. 
This probably should be very significant when the SCRPA eqs (\ref{S-term}) are solved with non restricted indices where the difference of occupation numbers can become very small. In that case those configurations become decoupled from the physically relevant space. It is the same as working with a non-orthonormal basis like, e.g., with the RGM or GCM, when one has to diagonalise the norm kernel and eliminate all configurations with vanishingly small eigen values \cite{RS}.

\vspace{1cm}

\noindent
\section{Self-Consistent RPA from the Coupled Cluster Wave Function}

\noindent
To be self-contained, in this section, we will re-derive SCRPA from a different perspective which will have interesting connection with TDDM and which will give some insight into which kind of ground state is implicitly used in STDDM* and/or STDDM.

\noindent
Formally the SCRPA equations have been written down several times in the past \cite{NPA628, NPA512}. They can be qualified as some sort of Hartree-Fock-Bogoliubov (HFB) equations for fermionic $ph$ pairs and they most of the time have been presented as a non-linear eigenvalue problem to be solved by iteration. SCRPA theory has recently known important new developments concerning its theoretical foundation \cite{Jemai2}. 
This stems from the fact that it was shown in that reference that the SUB2 coupled cluster wave function

\begin{eqnarray}
|Z\rangle &=& e^{\hat Z}|\mbox{HF}\rangle
\nonumber \\
\mbox{with}~~~~~\hat Z &=& \frac{1}{4}\sum_{p_1p_2h_1h_2}z_{p_1p_2h_1h_2}a^+_{p_1}a^+_{p_2}a_{h_1}a_{h_2}
\label{Z}
\end{eqnarray}

\noindent
is the vacuum to the following generalised RPA operator

\begin{eqnarray}
\tilde Q^+_{\nu} &=& \sum_{ph}[\tilde X^{\nu}_{ph}a^+_pa_h -\tilde Y^{\nu}_{ph}a^+_ha_p \nonumber \\
               &+& \frac{1}{2}\sum_{php_1p_2}\eta_{php_1p_2}a^+_{p_1}a_{p_2}a^+_ha_p \nonumber\\
&-& \frac{1}{2}\sum_{phh_1h_2}\eta_{h_1h_2ph}a^+_{h_1}a_{h_2}a^+_ha_p.
\label{Q+}
\end{eqnarray}

\noindent
That is there exists the killing condition

\begin{equation}
\tilde Q_{\nu}|Z\rangle = 0
\label{killing}
\end{equation}

\noindent
with the following relations between the various amplitudes 

\begin{eqnarray}
 \tilde Y^{\nu}_{ph} &=& \sum_{p'h'}z_{pp'hh'}\tilde X^{\nu}_{p'h'}\nonumber\\ 
z_{pp'hh'} &=& \sum_{\nu}\tilde Y^{\nu}_{ph}(\tilde X^{-1})^{\nu}_{p'h'}\nonumber\\
\eta^{\nu}_{p_1p_2ph} &=& \sum_{h_1}z_{pp_2hh_1}\tilde X^{\nu}_{p_1h_1}\nonumber\\
\eta^{\nu}_{h_1h_2ph}&=&\sum_{p_1}z_{pp_1hh_2}\tilde X^{\nu}_{p_1h_1}.
\label{amplitudes}
\end{eqnarray}

The amplitudes $z_{pp'hh'}$ are antisymmetric in $pp'$ and $hh'$. With the above relations, the vacuum state is entirely expressed by the RPA amplitudes $\tilde X, \tilde Y$. We remark that this vacuum state is exactly the one of coupled cluster theory (CCT) truncated at the two body level \cite{BR}. However, the use we will make of this vacuum is very different from CCT. Of course, for the moment, all remains formal because this generalized RPA operator contains, besides the standard one body terms, also specific two-body terms which cannot be handled in a straightforward way. For instance, this non-linear transformation cannot be inverted in a simple manner. However, we find the mere existence of an exact killing operator of the coupled cluster ground state quite remarkable.
 One may develop approximate methods to cope with those extra two-body terms. A first simple approximation consists in replacing in (\ref{Q+}) the occupation number operators in the $\eta$ terms by their expectation values, that is $a^+_{p_2}a_{p_1} \rightarrow \langle a^+_{p_1}a_{p_1}\rangle \delta_{p_1p_2}$ and $a^+_{h_1}a_{h_2} \rightarrow \langle a^+_{h_1}a_{h_1}\rangle \delta_{h_1h_2}$ where we supposed that we work in a basis where the single particle density matrix is diagonal. With the definition of the occupation numbers $n_k = \langle a^+_ka_k\rangle $, we then obtain the following approximate form of the $Q$-operator in (\ref{Q+})

\begin{eqnarray}
\tilde Q_{\nu} &=& \sum_{ph}[\tilde X^{\nu}_{ph} a^+_ha_p - \tilde Y^{\nu}_{ph}a^+_pa_h]\nonumber\\
&+&\frac{1}{2}\sum_{php_1}\eta_{p_1p_1ph}n_{p_1}a^+_pa_h \nonumber\\
&-& \frac{1}{2}\sum_{phh_1}\eta_{h_1h_1ph}n_{h_1}a^+_pa_h.
\label{Qkill}
\end{eqnarray}

\noindent
Evidently, this approximation, though suggestive, violates the killing condition (\ref{killing}). However, as has been shown in \cite{Jemai2}, the violation remains quite moderate. On the other hand, this approximation leads to a renormalisation of the $\tilde Y$ amplitudes in (\ref{Qkill}) and, therefore, we are back to the usual RPA operator with the one-body terms in (\ref{Qkill}) only. 
For simplicity, we will not change the nomenclature of the $\tilde Y$ amplitudes in the following. In spite of the approximation, 
we will henceforth assume that the killing condition still holds. However, we always should be aware that this only is true approximately with the atrophied form of the generalized RPA operator (\ref{Qkill}). The amplitudes $(\tilde X, \tilde Y)$ form a complete orthogonal set of vectors as explained, e.g., in \cite{RS}. We, therefore can invert the approximate RPA operator to obtain

\begin{equation}
a^+_pa_h = \sqrt{n_h-n_p}\sum_{\nu}[X^{\nu}_{ph}Q^+_{\nu} + Y^{\nu}_{ph}Q_{\nu}],
\label{inverse}
\end{equation}

\noindent
where we defined new amplitudes $X, Y$ via

\begin{equation}
\tilde X^{\nu}_{ph} = X^{\nu}_{ph}/\sqrt{n_h-n_p}; ~~~~~~\tilde Y^{\nu}_{ph} = Y^{\nu}_{ph}/\sqrt{n_h-n_p}
\label{tildeX}
\end{equation}

\noindent
and new RPA operators $Q_{\nu} = \sum_{ph}[ X^{\nu}_{ph} a^+_ha_p -  Y^{\nu}_{ph}a^+_pa_h]/\sqrt{n_h-n_p}$
\noindent
so that the state $|\nu\rangle = Q^+_{\nu}|Z\rangle$ is normalized, i.e., $\langle \nu|\nu\rangle = \langle Z|[Q_{\nu},Q^+_{\nu}]|Z\rangle/\langle Z|Z\rangle =1$ with

\begin{equation}
\sum_{ph}[|X^{\nu}_{ph}|^2 - |Y^{\nu}_{ph}|^2] = 1.
\label{XYnorm}
\end{equation}

The use of the CCT state $|Z\rangle$ has the great advantage that now in the calculation of the expectation values where we also need the occupation numbers expressed in terms of the $X, Y$ amplitudes, this can be achieved in a natural manner (this was in the past always a certain problem with SCRPA without the use of the CCT state). For example, we have

\begin{equation}
a^+_ha_h|Z\rangle = e^{\hat Z}\tilde J_{hh}|\mbox{HF}\rangle
\label{h+h}
\end{equation}

\noindent
with $J_{hh} = a^+_ha_h$ and $\tilde J_{hh} = e^{-\hat Z}J_{hh}e^{\hat Z} = J_{hh} + [J_{hh},\hat Z]$. Evaluating the commutator and then using the relation

\begin{equation}
\sum_{\nu}(\tilde X^{-1})^{\nu}_{p'h'}Q_{\nu} = a^+_{h'}a_{p'} - \sum_{ph}z_{pp'hh'}a^+_pa_h
\label{Qdestr}
\end{equation}

\noindent
we arrive at

\begin{eqnarray}
n_h &=& \langle a^+_ha_h\rangle \equiv \frac{\langle Z|a^+_ha_h|Z\rangle}{\langle Z|Z\rangle}\nonumber\\
&=&1-\frac{1}{2}\sum_p\langle a^+_pa_ha^+_ha_p\rangle \nonumber\\
&=& 1-\frac{1}{2}\sum_p[n_p\bar n_h -C_{phph}].
\label{nh}
\end{eqnarray}

\noindent
This relation can be used in (\ref{gs}) to have a fully closed system of equations. For the evaluation of the two-body term in terms of the $Y$-amplitudes, we will use the inversion of the $Q$-operators and obtain

\begin{equation}
n_h \equiv \langle a^+_ha_h\rangle = 1-\frac{1}{2}\sum_{p,\nu}(n_h-n_p)|Y^{\nu}_{ph}|^2.
\label{nh2}
\end{equation}

\noindent
The same can be repeated for $n_p$

\begin{eqnarray}
n_p \equiv \langle a^+_pa_p\rangle &=& \sum_h \frac{1}{2}\sum_h \langle a^+_pa_ha^+_ha_p\rangle \nonumber\\
&=&\frac{1}{2}\sum_h [n_p\bar n_h -C_{phph}] \nonumber\\
&=& \frac{1}{2}\sum_{h,\nu}(n_p-n_h)|Y^{\nu}_{ph}|^2 
\label{np}
\end{eqnarray}

\noindent
leading to a linear system of equations for $n_p, n_h$ which can be solved. The quadratic occupation number fluctuations can be treated in a similar way. They are related to $C_2$'s with either four particle or four hole indices. They can be approximated to leading order by quadratic forms of $C_2$'s with $pphh$ indices as shown in \cite{TS14}

\begin{equation}
C_{p_1p_2p_3p_4} \simeq \frac{1}{2}\sum_{hh'}C_{p_1p_2hh'}C_{hh'p_3p_4} + ...
\label{C4p}
\end{equation}

\begin{equation}
C_{h_1h_2h_3h_4} \simeq \frac{1}{2}\sum_{pp'}C_{h_1h_2pp'}C_{pp'h_3h_4} + ... .
\label{C4h}
\end{equation}

\noindent
We now can express all correlation functions and densities in $A$ and $B$ matrices by the RPA amplitudes $X, Y$ and, thus, have a fully self-consistent system of equations for $X, Y$. It should be mentioned, however, that due to the fact that the present RPA operator only contains $ph(hp)$ configurations, sum rules, Goldstone theorem, etc. are not strictly fullfilled. The violations usually remain very weak though, see \cite{Lipkin3}.\\

\noindent
There exists, however, a different closing of the equations employing the so-called selfconsistent particle-particle RPA (SCppRPA ~\cite{Hirsch}). 
It can be shown that the coupled cluster wave function is not only the vacuum to a generalised RPA operator in the $ph$ channel but also in the $pp(hh)$ channel. 
This is explained in Ref. \cite{Jemai2}. From SCppRPA one can naturally obtain the $C_2$'s with four particle or four hole indices, that is $C_{p_1p_2p_3p_4}$ and $C_{h_1h_2h_3h_4}$. Also the SCppRPA couples via the nonlinearity back to the particle-hole SCRPA considered here \cite{NPA628}.\\

Iterating SCphRPA and SCppRPA simultaneously, again corresponds approximately to summing the parquet diagrams already discussed above.\\

\vspace{0.5cm}
\section{   Short description of connection of SCRPA with Green's functions}

In condensed matter physics dealing with homogeneous infinite systems, one usually does not formulate the problems in the form of an eigenvalue equation. One rather employs propagators or many body Green's functions. Of course, it is clear that every eigenvalue problem has a corresponding formulation with Green's functions but it may be useful to give some more  details on the ingredients of the present formalism. The Green's function equivalent to the eigenvalue equation of SCRPA 
(\ref{scrpa}, \ref{SCRPA}) is, in a way,  somewhat particular. As one may immediately realise, it cannot come from the familiar many time Green's function approach where, e.g., the two body propagator (and also its integral kernel) depends on four times once one goes beyond the standard HF-RPA scheme. This stems from the fact that in an eigenvalue problem only one energy (the eigenvalue) is involved and then the corresponding integral equation for the Green's function also can involve only one energy, even in the integral kernel.  
Though the formalism has been described in earlier publications, see, for instance, refs. \cite{NPA628}\cite{Storo},
we feel that it may be helpful for the reader to give a short outline of the procedure. 
To this purpose, we write down the corresponding integral equation form of (\ref{scrpa}), that is the Bethe-Salpeter equation

\begin{eqnarray}
&(& \omega - E_{k_1} +E_{k_2})\tilde{\mathcal G}^{\omega}_{k_{1}k_2k_{3}k_4} \nonumber\\
&=& {\mathcal N}_{0, k_1k_2}[
\delta_{k_1k_3'}\delta_{k_2k_4'} +\sum_{k_{3'}k_{4'}}   {\mathcal S}_{k_1k_{2}k_{3'}k_{4'}}]
\nonumber \\
&\times&\tilde{\mathcal G}^{\omega}_{k_{3'}k_{4'}k_{3}k_4}.
\label{BSE}
\end{eqnarray}

Inserting the spectral representation of the Green's function

\begin{equation}
\tilde{\mathcal G}^{\omega} = \sum_{\nu}\frac{\chi^{\nu}N_{\nu}\chi^{\nu^*}}{ \omega - \Omega_{\nu} + i\eta N_{\nu}}
\label{spectralGF}
\end{equation}
where the sum goes over positive and negative values of $\nu$ and $N_{\nu} = - N_{-\nu}, \Omega_{\nu} = -\Omega_{-\nu}$,
and taking the limit $ \omega \rightarrow \Omega_{\nu}$, we obtain in comparing the singularities on left and right hand sides, the eigenvalue equation (\ref{scrpa}). 

In order to see how this scheme with the equation of motion technique can go on and lead to an $\omega$-dependent term in the integral kernel of the Bethe-Salpeter equation, we consider the operator (\ref{Q3}) to include a two body term as a first extension, eventually higher order terms.

Eliminating the 2-body amplitudes from the coupled equations of one body and two body amplitudes, one obtains an effective equation for the $\chi$ amplitudes with an effective, energy dependent potential containing implicitly the two body amplitudes. This effective potential can be qualified to corresponds to the $\omega$ dependent part of a two body self energy. This procedure can formally be pushed up to the N-body amplitudes leading thus to an exact two body equation of a Dyson equation form in analogy to what is known from the single particle Green's function.

Let us shortly show how the same scheme can be obtained beginning directly with the Green's function. We start with the following chronological propagator

\begin{equation}
{\mathcal G}_{12}^{t-t'} = -i \langle 0|\mbox{T}A_1(t)A_{2}^+(t')|0\rangle,
\label{defG}
\end{equation}

\noindent
with $A(t) = e^{iHt}A(0)e^{-iHt}$, T the time ordering operator and

\[A_1 = a^{+}_{k_{1'}}a_{k_1}~,~~~~ A_{2}^{+}= a^{+}_{k_2}a_{k_{2'}} \]

\noindent
where $a^+, a$ are fermion creation and destruction operators, respectively and the Green's function in (\ref{defG}) is thus a density-density correlation function. It is always understood that the indices $k_i$ comprise, as before, momentum and spin and, eventually more quantum numbers, such as isospin, etc. We remark that in this definition of the Green's function we put pairs of fermion operators on equal times so that the Green's function depends only on one time difference at equlibrium. 
The $\tilde {\mathcal G}$ function is related to ${\mathcal G}$ in replacing in the latter the $A_1$ by $\tilde A_1 = a^{+}_{k_{1'}}a_{k_1}/\sqrt{N_{k_{1'}k_1}}$, etc.
We now claim that for this two time Green's function, one can write down in a well defined way a formally exact integral equation with an integral kernel which also depends only on one time difference (or in energy space on one energy $ \omega$). We, thus, write 

\begin{equation}
{\mathcal G}^{\omega} = {\mathcal G}^{\omega}_0 + {\mathcal G}^{\omega}_0\Sigma^{\omega}
{\mathcal G}^{\omega}~,
\label{BSGF}
\end{equation}

\noindent
where it is understood that this is a matrix equation with matrix multiplication of the various products. The lowest order Green's function ${\mathcal G}_0$ is thereby given for, e.g., a translationally invariant system as

\begin{equation}
{\mathcal G}_{0,12}^{\omega} =\frac{n_{k'_1} - n_{k_1}}{ \omega -E_{k_1} + E_{k'_1}}\delta_{k_1k_2}\delta_{k'_1k'_2}
\label{G_0}
\end{equation}

\noindent
where $n_k = \langle 0|a^{+}_ka_k|0\rangle$ are the single particle occupation numbers and $E_k = k^2/(2m) + \sum_{k'}\bar v_{kk'kk'}n_{k'}$ are the mean field energies.

In principle, Eq. (\ref{BSGF}) may thus serve as a definition of the kernel $\Sigma^{\omega}$. It turns out that $\Sigma^{\omega}$ is a well defined object for which expressions in terms of usual correlation functions and Green's functions can be given, see, e.g., \cite{NPA628}. This kernel can be considered as some kind of higher order self energy, here the self-energy of density fluctuations. As the well known self-energy of the single particle Green's function, it splits into an instantaneous, energy independent part $\Sigma^0$ and an explicitly energy dependent part $\Sigma^r(\omega)$. It can be shown that $\Sigma^0$ is equivalent to the matrix ${\mathcal S}$ in (\ref{scrpa}) as this is explained in \cite{NPA628}. Therefore (\ref{scrpa}) and (\ref{BSGF})) are equivalent once $\Sigma^{\omega}$ is replaced by its static part $\Sigma^0$. Mathematically, this can be seen quite straightforwardly in applying the equation of motion to the propagator (\ref{defG}): $i\frac{\partial}{\partial t}{\mathcal G}_{12} = \delta(t-t')\langle 0|[A_1,A^+_2]|0\rangle -i\langle 0|\mbox{T}[A_1,H]_tA^+_2(t')|0\rangle$. Applying now the equation of motion a second time to the time $t'$ figuring in the correlation function which appears on the r.h.s. of this equation, one realises that the part which acts on the chronological operator $\mbox{T}$ leads to the double commutator also involved in ${\mathcal S}$ of Eq.(\ref{scrpa}) and, consequently, in the instantaneous part of the self energy $\Sigma^0$. The application of the time-derivative on $t'$ contained in $A^+_2(t')$ will lead to the energy dependent part of the self-energy in (\ref{BSGF}). This brief outline should only serve to give the reader a quick feeling how such a somewhat unusual integral equation like (\ref{BSGF}) with an integral kernel depending only on one energy can be obtained. For a more detailed outline, we refer the reader to \cite{NPA628}. \\

Concerning the practical solution of (\ref{BSE}), it can be seen
 from (\ref{scrpa}), that the static part only contains up to two body correlation functions which can be calculated from (\ref{BSE}) and, thus, a self-consistent cycle is established. As just explained, the dynamic, explicitly energy dependent part contains the coupling to higher configurations involving four body propagators. Their inclusion leads in some approximation to what is known in the equation of motion method as the second RPA equations ~\cite{2ndrpa}.\\

It may be worth mentioning that a perturbative analysis of $\Sigma$ in (\ref{BSGF})) shows that the terms are not equivalent to Feynman diagrams. Nevertheless, one can present the various terms in $\Sigma^0$ (or equivalently in ${\mathcal S}$ of eq (\ref{scrpa})) by the graphs shown in Fig. \ref{scrpa-graph1}. If in this figure the two body correlation functions are replaced by the first oder expression in the interaction, the standard second order perturbation graphs emerge with, however, the particularity that they occur instantaneously, that is they do not propagate. Even, if the correlation functions in Fig. \ref{scrpa-graph1} are replaced by their full expression, the graphs stay, as indicated in the figure, instantaneous. This feature results from the minimisation of the energy weighted sum rule as explained in section 2.\\

Similar type of equations with integral-kernels depending only on one frequency are obtained from the hypernetted chain equations, see \cite{Saar}.

\vspace{2cm}

\section{Applications}
\subsection{Preliminaries}

In order to guide the reader in the following applications with the various approximations used, let us make a short summary here.\\

First, there is the TDDM method, described in Sect.II. It allows to calculate the occupation numbers $n_k$ and the four types of 2-body correlation functions considered. Disposing of those quantities allows to calculate the total ground state energy or various partial quantities thereof, as. e.g., the so-called 2-body correlation energy. The $n_k$ and $C_2$'s can also be used to set up the correlated RPA matrix, in which case we talk about the C-RPA scheme. The C-RPA and SCRPA schemes appear naturally as the one body sector of the linearised TDDM equations. The latter equations have been called either STDDM*-b or STDDM-b equations according to whether one includes the approximate form of the 3-body correlation function $C_3$, Eq. (\ref{3body2}), or not. Let us recall that the one body sector of STDDM-b and STDDM*-b is not affected by $C_3$ when the 2-body space is decoupled from the 1-body one. The non-linearity in $C_2$'s only affects the 2-body sector as seen when comparing (\ref{SCS'}) with (\ref{SCS''}). There also exist STDDM and STDDM* equations which are equivalent but very non-symmetric versions of STDDM-b and STDDM*-b. They are not considered in the applications. Finally there exists the so-called Extended Second RPA (ESRPA) equation which does not follow from the TDDM approach but is obtained from a minimisation of the energy weighted sum-rule involving 1-body and 2-body operators. Since STDDM-b and STDDM*-b equations can be shown to be  approximate 
forms of ESRPA, we consider ESRPA (slightly) superior to all the other kinds of equations we have established. One should realise, however, that STDDM-b, STDDM*-b, and ESRPA which all include the 2-body sector can be solved for the model cases presented below which involve limited configuration spaces but for realistic problems as the homogeneous electron gas or nuclear matter, etc., one must be happy if the equations of the 1-body sector, that is C-RPA and/or SCRPA can be tackled. One should appreciate the following results in the light of these preliminary remarks.

\subsection{Lipkin model}
We first consider the Lipkin model \cite{Lip}.
The Lipkin model describes an $N$-fermions system with two
$N$-fold degenerate levels with energies $\epsilon/2$ and $-\epsilon/2$,
respectively. The upper and lower levels are labeled by quantum number
$p$ and $-p$, respectively, with $p=1,2,...,N$. We consider
the standard Hamiltonian
\begin{equation}
\hat{H}=\epsilon \hat{J}_{z}+\frac{V}{2}(\hat{J}_+^2+\hat{J}_-^2),
\label{elipkin}
\end{equation}
where the operators are given as
\begin{eqnarray}
\hat{J}_z&=&\frac{1}{2}\sum_{p=1}^N(a_p^{+}a_p-{a_{-p}}^{+}a_{-p}), 
\label{elipkin1}\\
\hat{J}_{+}&=&\hat{J}_{-}^{+}=\sum_{p=1}^N a_p^{+}a_{-p}.
\label{elipkin2}
\end{eqnarray}
The operators $J_z,J_{\pm}$ are pseudospin operators and fullfill commutation relations of angular momenta.

\noindent
\begin{figure}
\resizebox{0.5\textwidth}{!}{%
\includegraphics{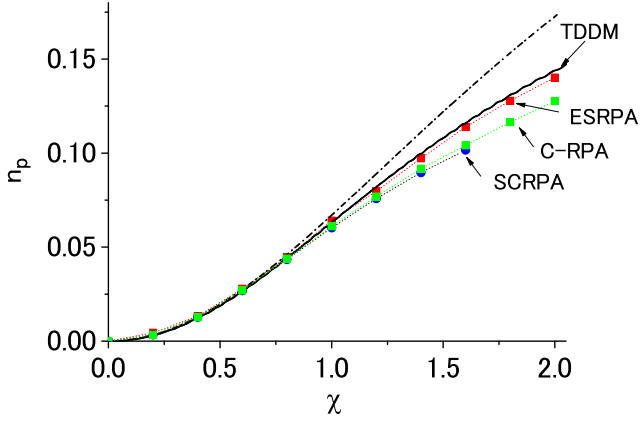}
}
\caption{Occupation probabilities of the upper state calculated in TDDM (solid line), ESRPA (red squares), C-RPA (green squares)
and SCRPA (blue circles)
as a function of $\chi=(N-1)|V|/\epsilon$ for $N=4$. The exact solution is shown with the dot-dashed line.
The occupation probability and correlation
matrix in TDDM are used in the C-RPA and ESRPA calculations.}
\label{lip1}
\end{figure}

\begin{figure}
\resizebox{0.5\textwidth}{!}{%
\includegraphics{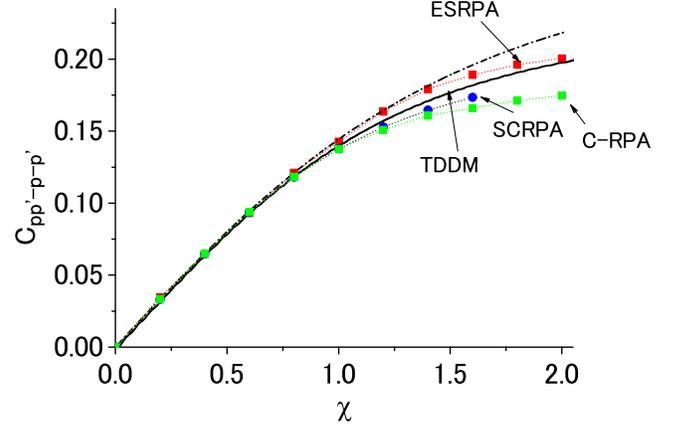}
}
\caption{Same as Fig. \ref{lip1} but for the two-body correlation matrix $C_{pp'-p-p'}$.}
\label{lip2}
\end{figure}

\begin{figure}
\resizebox{0.5\textwidth}{!}{%
 \includegraphics{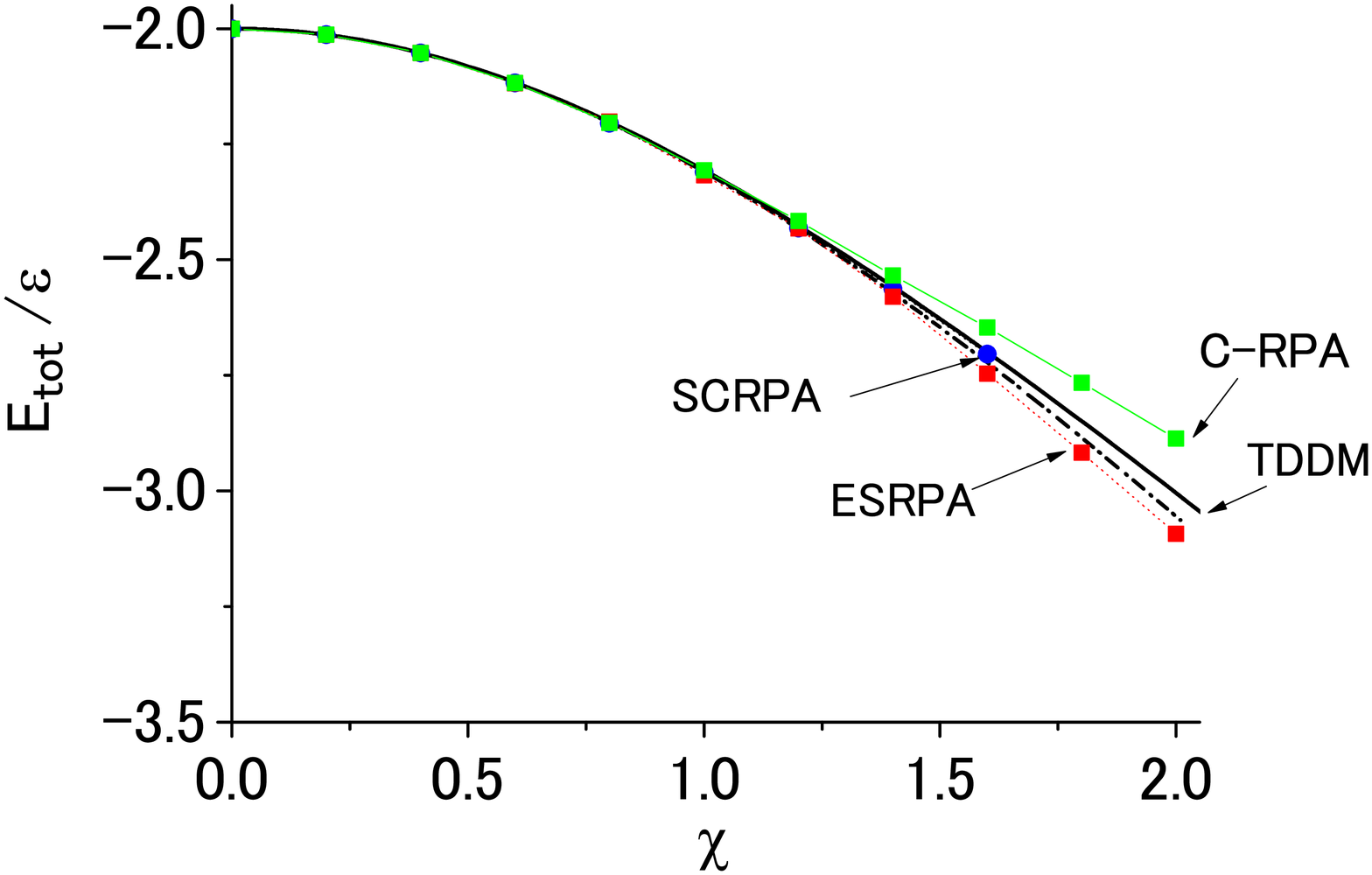}
}

\caption{Same as Fig. \ref{lip1} but for the ground-state energy.}
\label{lip3}
\end{figure}

\begin{figure}
\resizebox{0.5\textwidth}{!}{%
\includegraphics{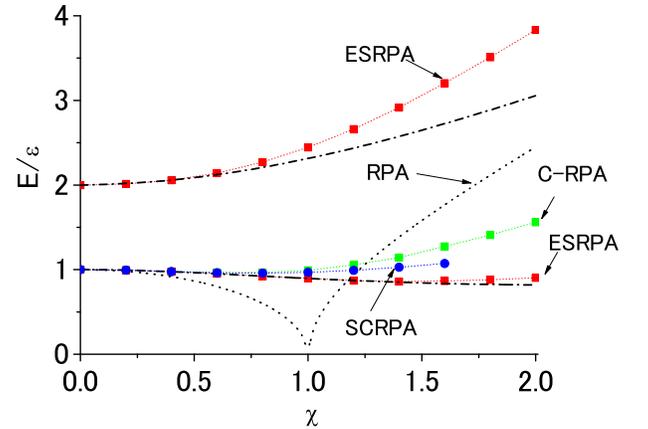}
}
\caption{Same as Fig. \ref{lip1} but for the excitation energies of the first and second excited states.
}
\label{lip4}
\end{figure}

\begin{figure}
\resizebox{0.5\textwidth}{!}{%
\includegraphics{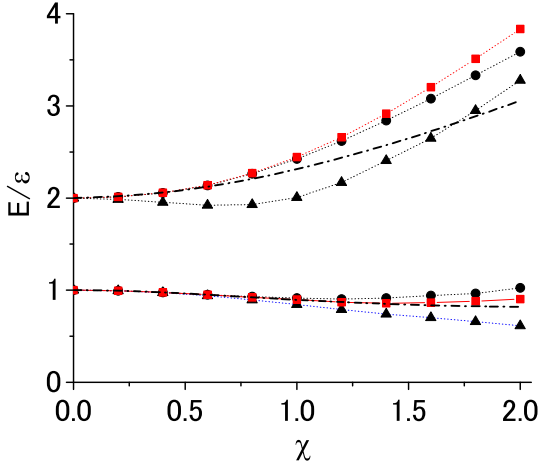}
}
\caption{Excitation energies of the first and second excited states calculated in STDDM-b (triangles), STDDM$^*$-b (circles)
and ESRPA (squares) as a function of $\chi=(N-1)|V|/\epsilon$ for $N=4$. The exact solution is shown with the dot-dashed line.}
\label{lipstddm}
\end{figure}

The ground state in TDDM is obtained using the
adiabatic method: Starting from the HF ground state, we solve the TDDM equations 
(Eqs. (\ref{n}) and (\ref{N3C2})) by gradually increasing the
residual interaction such that $V'=V\times t/T$, as described in section II.D. We use $T=4\times 2\pi/\epsilon$. 
For the 3-body terms in Eq. (\ref{N3C2}) we use the approximations Eqs. (\ref{C_2C_2-h}) and (\ref{C_2C_2-p}) which are supposed to be the leading terms. All possible single particle indices are taken into account one by one (the so-called m-scheme, see also \cite{Jansen}). The original basis is kept.\\

 In a first application, the occupation numbers $n_{\alpha}$ and 2-body correlation functions
$C_{\alpha\beta\alpha'\beta'}$ are determined from  the TDDM calculation and the RPA matrix is set up with these values. We refer to this scheme as the correlated RPA (C-RPA), see Sect.IV.A to 
distinguish it from SCRPA which takes into account self-consistency.
We found 
it  necessary to include the factor $1/2$ in Eq. (\ref{nh2}) when we consider non-collective amplitudes as $\chi^\mu_{-p',p}$ and $\chi^\mu_{p,-p'}$ in addition 
to $\chi^\mu_{-p,p}$ and $\chi^\mu_{p,-p}$, that is all possible RPA-amplitudes. 
When we keep only the collective amplitudes, the results deteriorate and in addition the factor 1/2 (\ref{nh2}) has to be suppressed. 
This is in line with the discussion about the factor 1/2 in the occupation number expressions by Rowe in \cite{Rowe} given a long time ago.\\
In a second application  we also performed self-consistent RPA calculations corresponding to Eq. (\ref{scrpa}), taking again all kinds of amplitudes, collective and non-collective, that is, we also included all the amplitudes $\chi^\mu_{-p',p}$ 
and $\chi^\mu_{p,-p'}$
and consequently the factor $1/2$ in Eq. (\ref{nh2}) was kept. In SCRPA the two-body correlation matrices $C_{p_1p_2p_3p_4}$ and $C_{h_1h_2h_3h_4}$
which are not directly related to the one-body transition amplitudes ($X, Y$) are calculated using Eqs. (\ref{C4p}) and 
(\ref{C4h}).
To calculate the $2p-2h$ elements figuring in the above expressions for $C_{pp'-p-p'}$ of the two-body correlation matrix, we use their relation with the RPA amplitudes given in 
 Eq. (\ref{R,K}) with Eq. (\ref{DM}).
The occupation probability $n_p$ of the upper state and the two-body correlation matrix
$C_{pp'-p-p'}$ calculated in TDDM (solid line) and ESRPA (red squares) are shown in Figs. \ref{lip1} and \ref{lip2}, respectively,
as a function of $\chi=(N-1)|V|/\epsilon$ for $N=4$. The RPA solution becomes unstable at $\chi=1$ as shown below in Fig.\ref{lip4}.
The results of SCRPA (round dots) are shown up to $\chi=1.6$ because beyond $\chi\approx 1.6$ the numerical solution becomes unstable.

The Lipkin model is simple enough to solve the complicated self-consistent ESRPA equations (\ref{SCS}), 
however still some approximations have been applied. For the three body correlation functions, again the approximations Eqs. (\ref{C_2C_2-h}) and (\ref{C_2C_2-p})  are employed.
The 4-body correlation functions $C_4$ contained in the $\cal{D}$-matrix are neglected. Furthermore, in the ESRPA calculations we included only the one-body amplitudes with the same quantum number 
( this corresponds to the collective subspace as usually considered in RPA) such as
$\chi^\mu_{-p,p}$, $\chi^\mu_{p,-p}$, $\chi^\mu_{-p,-p}$ and $\chi^\mu_{p,p}$ and used Eq. (\ref{ninSTDDM}) to obtain $n_{p}$.\\
All two body amplitudes ${\mathcal X}_{\alpha \beta \alpha' \beta'}$ with either $p_{\alpha} = p_{\alpha'}; p_{\beta} = p_{\beta'}$ or $p_{\alpha} = p_{\beta'}; p_{\beta} = p_{\alpha'}$ are included,
where $p_\alpha$ is the $p$ quantum number given in Eqs. (\ref{elipkin1}) and (\ref{elipkin2}).\\

 The occupation numbers and 2-body correlation functions shown in Figs. \ref{lip1} and \ref{lip2} are very sensitive quantities concerning the underlying wave function. Let us mention again that it is important for the accuracy of the results to work with all possible amplitudes (collective and non-collective), that is with the m-scheme. Taking into account only collective amplitudes sensitively deteriorates the results (not shown in the figures).
SCRPA and C-RPA are about  on same grounds, since they both work with the m-scheme and take the non-linearities in the $C_2$'s into account. ESRPA and TDDM are also more or less equivalent, since they both take into account two body amplitudes, see section III.C. We may, however, remark that in realistic situations ESRPA may be inapplicable, besides in very restricted configuration spaces, because of its numerical complexity whereas this is not the case with TDDM. As a general remark, we can say that 
all approximations perform quite well up to $\chi = 1$ but start to deviate more or less strongly from the exact result (dot dashed line) thereafter. SCRPA and C-RPA are simpler than the approaches including the two body sector because the dimensions of the matrices remain much smaller in the first case. 
The value $\chi = 1$ is the one where standard RPA becomes unstable and a change of the single particle basis becomes necessary ( the 'deformed' basis). 
Here, we do not operate a change of basis but still the system seems to feel the entering into a new 'phase'. 
We should also remember that  $N$ = 4, is the worst case where the quantum fluctuations are the strongest (the $N$ =2 case being more or less trivial becomes exact in SCRPA), see \cite{Jemai} (anticipating, this will also be the case in the other two models treated below). The results will improve for higher values of $N$.

The ground-state energies in TDDM (solid line), ESRPA (red squares), C-RPA (green squares) and SCRPA (blue circles) are shown
in Fig. \ref{lip3} as a function of $\chi$ for $N=4$. The exact values are again given with the dot-dashed line.
The ground state energy in ESRPA is calculated using $n_p$ and $C_{pp'-p-p'}$ given in Figs. \ref{lip1} and \ref{lip2}.
All calculations agree well with the exact values. The ground state energy is a more robust quantity than are, e.g., the occupation numbers.

The excitation energies of the first and second excited states are displayed in Fig. \ref{lip4} as a function of $\chi$. We see that ESRPA performs extremely well, even far beyond the RPA instability point of $\chi = 1$. C-RPA and SCRPA also are very good but deteriorate after the instability point. Apparently the selfconsistency (SCRPA) brings, in the domain where the results are stable, a slight advantage over the non-selfconsistent one (C-RPA) but this may not be very significant in general cases.
In the case of the second excited state which can be obtained with ESRPA, deviation from the exact solution 
becomes larger with increasing $\chi$. This can be  explained either by the neglect of the coupling to higher amplitudes or by the fact that in ESRPA non-collective amplitudes are not included. 
Let us mention again that ESRPA can only be tackled at the moment for simple models. In realistic cases this approach becomes numerically too complicated.\\
The excitation energies of the first and second excited states calculated in STDDM-b (triangles), STDDM$^*$-b (circles)
and ESRPA (squares) are shown in Fig. \ref{lipstddm} as a function of $\chi=(N-1)|V|/\epsilon$ for $N=4$. The exact solution is shown with the dot-dashed line. Figure \ref{lipstddm} shows that STDDM$^*$-b is a good approximation
to ESRPA but up to $\chi =1$, STDDM-b also works quite well. All two body amplitudes have been taken into account, that is
${\mathcal X}_{pp'-p-p'}$, ${\mathcal X}_{-p-p'pp'}$,
${\mathcal X}_{p-p'p-p'}$, ${\mathcal X}_{pp'pp'}$, ${\mathcal X}_{-p-p'-p-p'}$, ${\mathcal X}_{p-p'-p-p'}$, ${\mathcal X}_{-p-p'-pp'}$,
${\mathcal X}_{pp'p-p'}$, and ${\mathcal X}_{-pp'pp'}$.

Let us remind that the difference between ESRPA and STDDM-b and STDDM*-b is that in STDDM $C_3$ is totally neglected. 
The small difference between STDDM*-b and ESRPA originates in the fact that ${\cal D}$ in Eq. (\ref{SCS}) is
not the same as $c{\mathcal T}+\tilde d{\mathcal N}_2$ in Eq. (\ref{SCS''}).\\

\begin{figure}
\resizebox{0.5\textwidth}{!}{%
\includegraphics{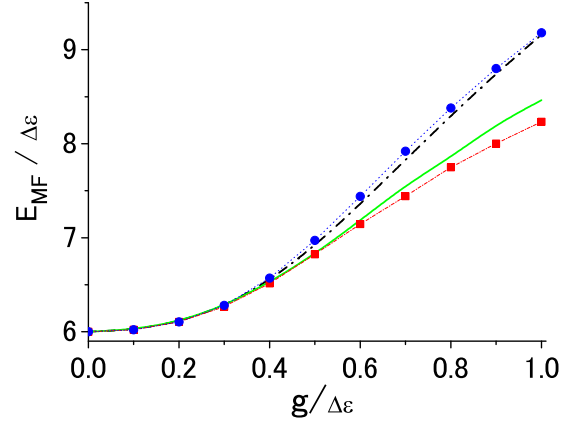}
}
\caption{Mean-field energy $E_{\rm MF}$ calculated in ESRPA (squares) and SCRPA (circles)
as a function of $g/\Delta \epsilon$ for $\Omega =N=6$. The TDDM results and the exact values are shown
with the solid and dot-dashed lines, respectively.}
\label{pickMF}
\end{figure}

\begin{figure}
\resizebox{0.5\textwidth}{!}{%
 \includegraphics{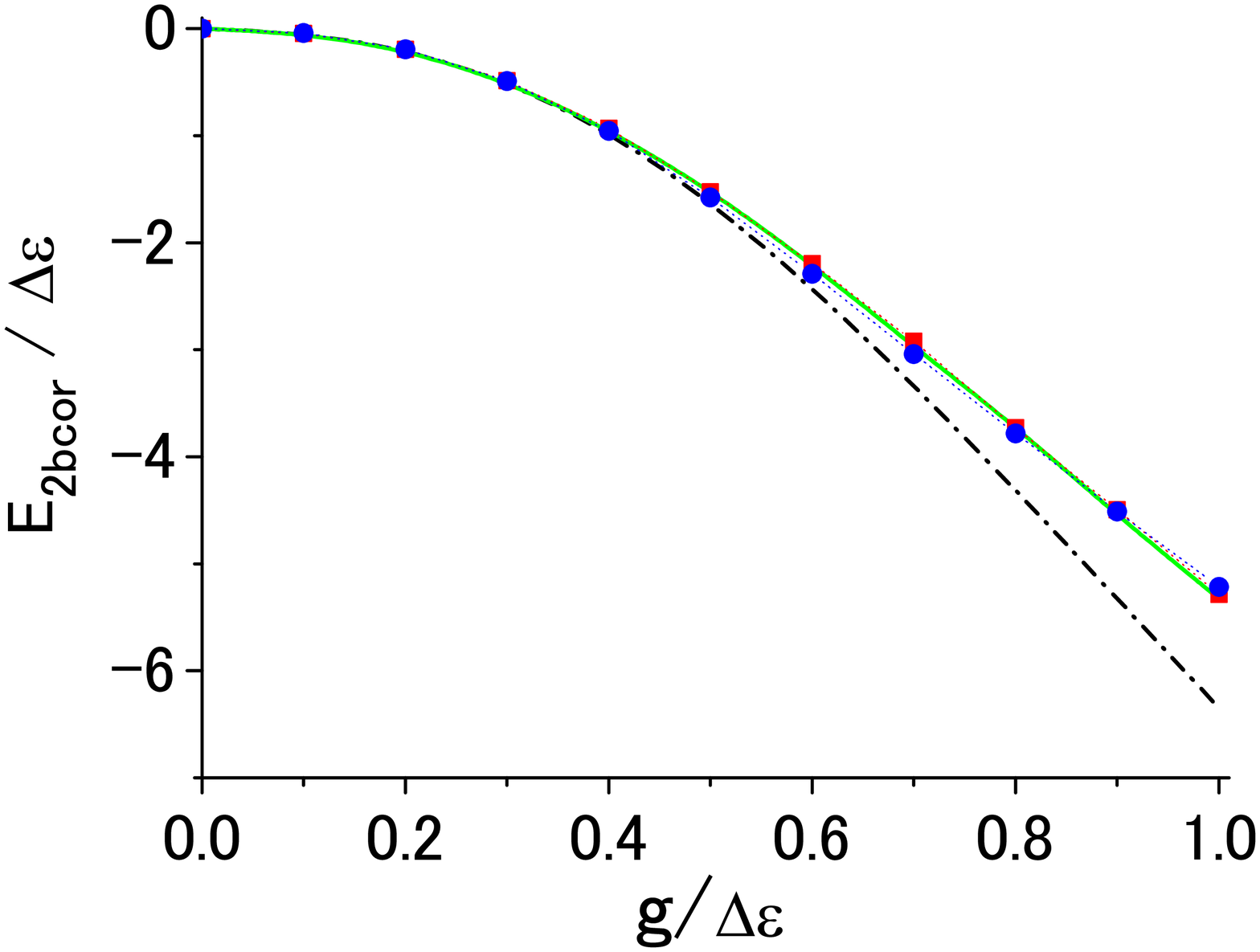}
}
\caption{Same as Fig. \ref{pickMF} but for the 2-body correlation energy $E_{\rm 2bcor}$.}
\label{pickEC}
\end{figure}

\begin{figure}
\resizebox{0.5\textwidth}{!}{%
\includegraphics{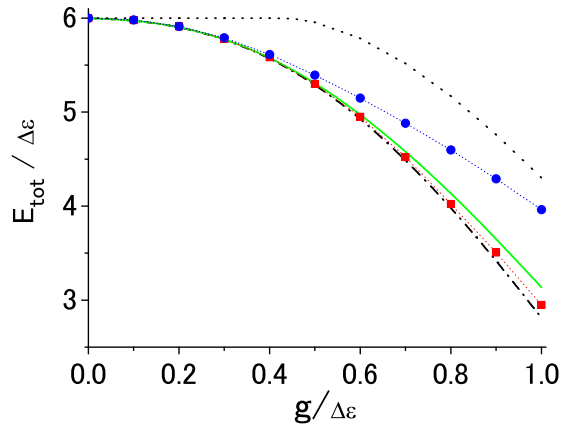}
}
\caption{Same as Fig. \ref{pickMF} but for the ground state energy $E_{\rm tot}$. The dotted line dipicts the results in BCS.}
\label{pickE}
\end{figure}

\subsection{Pairing model}
Next we consider the 
pairing Hamiltonian \cite{rich}
\begin{eqnarray}
H=\sum_{i=1}^\Omega\epsilon_\alpha(a^+_i a_i+a^+_{\bar{i}}a_{\bar{i}})
-g\sum_{i\neq j}^\Omega a^+_i a^+_{\bar{i}}a_{\bar{j}}a_j.
\label{hpair}
\end{eqnarray}
Here $g$ is the strength of the pairing force acting in a space of $\Omega$ twofold degenerate equidistant orbitals
with the single-particle energies $\epsilon_i=(i-1)\Delta\epsilon$. This Hamiltonian has extensively been used to 
investigate the validity of theoretical approaches \cite{rich}. 

The ground state in TDDM is obtained using again the
adiabatic method. We use $T=6\times 2\pi/\Delta\epsilon$. 
Since there are several occupation probabilities and the number of the elements of the two-body correlation matrix is not small in the 
pairing Hamiltonian, we discuss their average properties using the correlation energy $E_{\rm 2bcor}$ (Eq. (\ref{ecorr})) and the mean-field energy $E_{\rm MF}$, which is
given by $E_{\rm MF}=\sum_{\alpha}\epsilon_\alpha n_{\alpha\alpha}$ in the case of the pairing model.
The mean-field energy in ESRPA is calculated from the occupation probabilities given by Eq. (\ref{ninSTDDM})  
and the correlation energy in ESRPA given by Eq. (\ref{Eint3}). 
Since the first term on the right-hand side of Eq. (\ref{Eint4}) does not exist in the pairing Hamiltonian Eq. (\ref{hpair}),
Eq. (\ref{Eint3}) holds in ESRPA. \\

 In principle SCRPA is not adequate for the solution of the pairing case because it is essentially a particle-hole theory. Nevertheless, SCRPA also includes some particle-particle correlations and  it is interesting to see how well the ph-SCRPA performs. In principle, however, one should better use the pp-SCRPA as described in Section V and applied in \cite{Storo}.
As before, in SCRPA all $p-h$ and $h-p$ amplitudes are taken and the factor 1/2 in Eq. (\ref{nh2}) is kept. 
The matrix elements $C_{p_1p_2p_3p_4}$ and $C_{h_1h_2h_3h_4}$ are calculated using Eqs. (\ref{C4p}) and (\ref{C4h}) in SCRPA.

The results in ESRPA (squares)
are compared with the results of other calculations in Figs. \ref{pickMF}--\ref{pickE} as a function $g/\Delta \epsilon$ for $\Omega=N=6$.
The results in SCRPA are given with the circles and those in TDDM with the solid line. The dot-dashed line depicts the exact values.
The mean-field energies in SCRPA are closer to the exact values than those in ESRPA, whereas
ESRPA and SCRPA give similar results as with TDDM for $E_{\rm 2bcor}$. The good agreement of SCRPA results with exact ones for the mean field energies may be an accident.
  
The results of TDDM and ESRPA agree with each other and for $E_{\rm tot}$ they are close to the exact values. 
The dotted line in Fig. \ref{pickE} 
depicts the results of BCS, which are in poor agreement with the exact solution. One also can read off the critical coupling strength $g \sim 0.43$.
Since $E_{\rm 2bcor}$ in SCRPA is not large enough to compensate large $E_{\rm MF}$, $E_{\rm tot}$ in SCRPA deviates from the exact values with increasing $g$.
It is easy to understand that SCRPA cannot give sufficient $E_{\rm 2bcor}$ in the pairing model: As said before, the two-body interaction in Eq. (\ref{hpair}) 
consists of $p-p$ and $h-h$ correlations
which cannot be fully included by 1p-1h excitation modes in SCRPA. There are only non-collective 1p-1h excitation modes in the case of the pairing model.
The good agreement of $E_{\rm MF}$ in SCRPA with the exact solution suggests that such non-collective 1p-1h excitation modes are well described by SCRPA.
In fact the excitation energy $E_1$ of the first excited state calculated in SCRPA is $E_1/\Delta\epsilon=3.44$ at $g/\Delta\epsilon=1$  and the
corresponding exact value is $3.54$. 
As mentioned above, $E_{\rm MF}$ and $E_{\rm 2bcor}$ in ESRPA are determined by the properties of two-phonon states. Therefore, deviations of the results in ESRPA from
the exact values indicate that description of the two-phonon states in ESRPA without coupling to higher amplitudes becomes not good with increasing interaction strength,
as is the case of the Lipkin model: The excitation energy $E_2$ 
of the first two-phonon state in ESRPA is $E_2/\Delta\epsilon=5.43$ at $g/\Delta\epsilon=1$, which is about 25 $\%$ larger than the exact value $4.37$. \\

All in all, one must say that the phSCRPA performs surprisingly well, at least up to the critical value g $\sim$ 0.43. So, it may be important in general to include ph correlations also in the pairing case. We have seen that ppSCRPA contains ph correlations and phSCRPA pp correlations. It may be a good idea to couple both channels in a self consistent approach. As discussed earlier, this self consistent coupling of $pp$ and $ph$ channels has some similarity with parquet diagram summation.

\begin{figure}
\resizebox{0.5\textwidth}{!}{%
\includegraphics{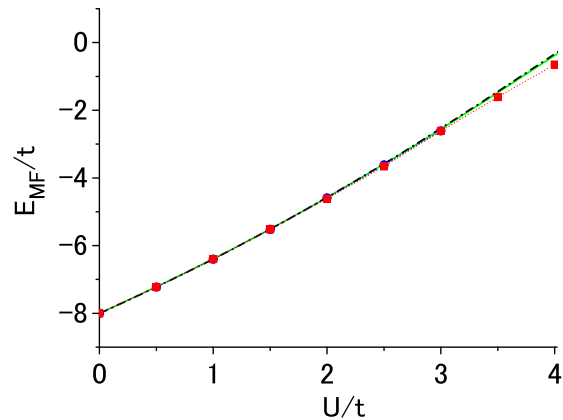}
}
\caption{Mean-field energy $E_{\rm MF}$ calculated in ESRPA (squares) and SCRPA (circles)
as a function of $U/t$ for the six-site Hubbard model with half-filling. The TDDM results and the exact values are shown
with the solid and dot-dashed lines, respectively.
}
\label{hubMF}
\end{figure}

\begin{figure}
\resizebox{0.5\textwidth}{!}{%
 \includegraphics{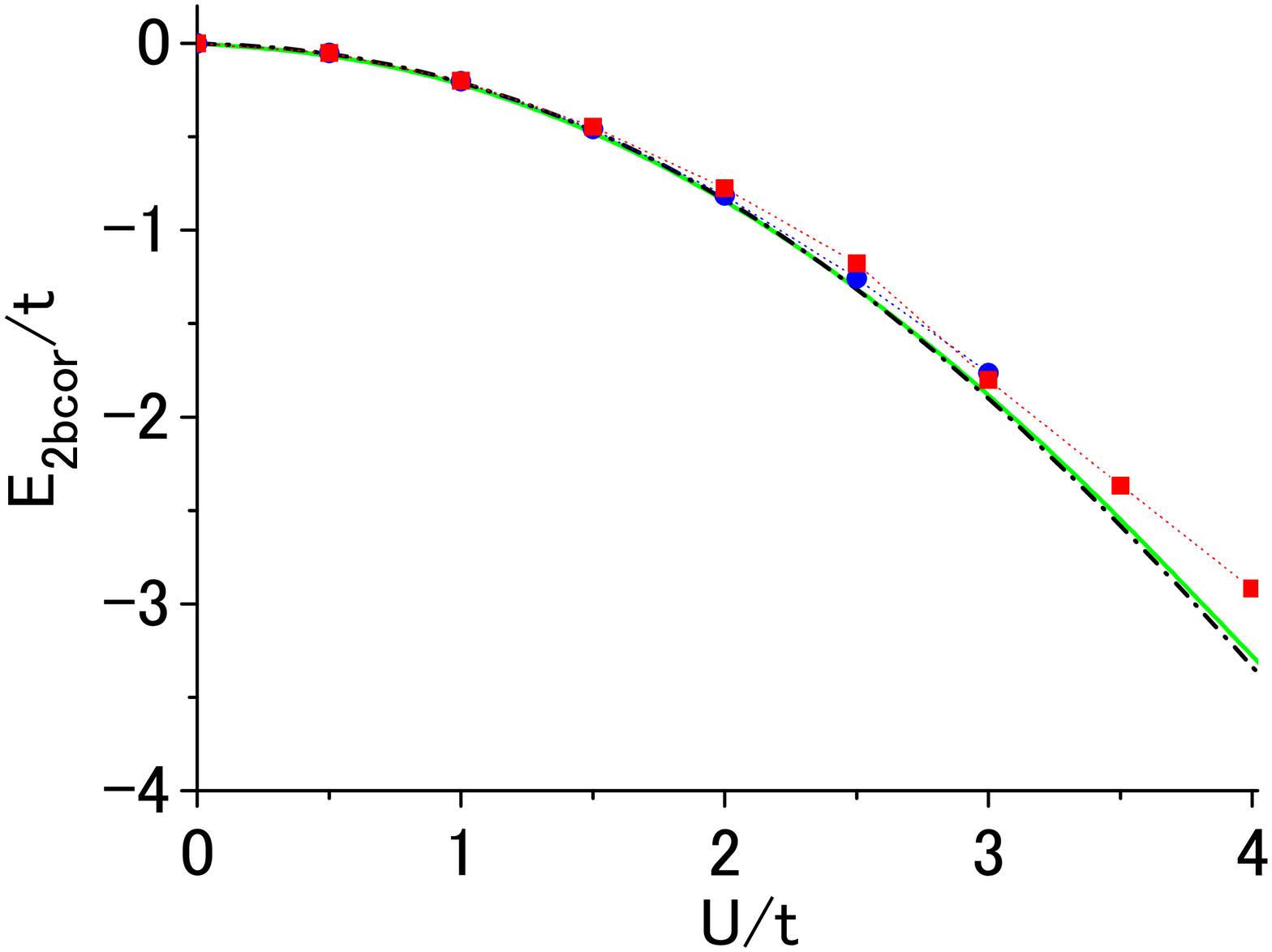}
}
\caption{Same as Fig. \ref{hubMF} but for the 2-body correlation energy $E_{\rm 2bcor}$.}
\label{hubEC}
\end{figure}

\begin{figure}
\resizebox{0.5\textwidth}{!}{%
\includegraphics{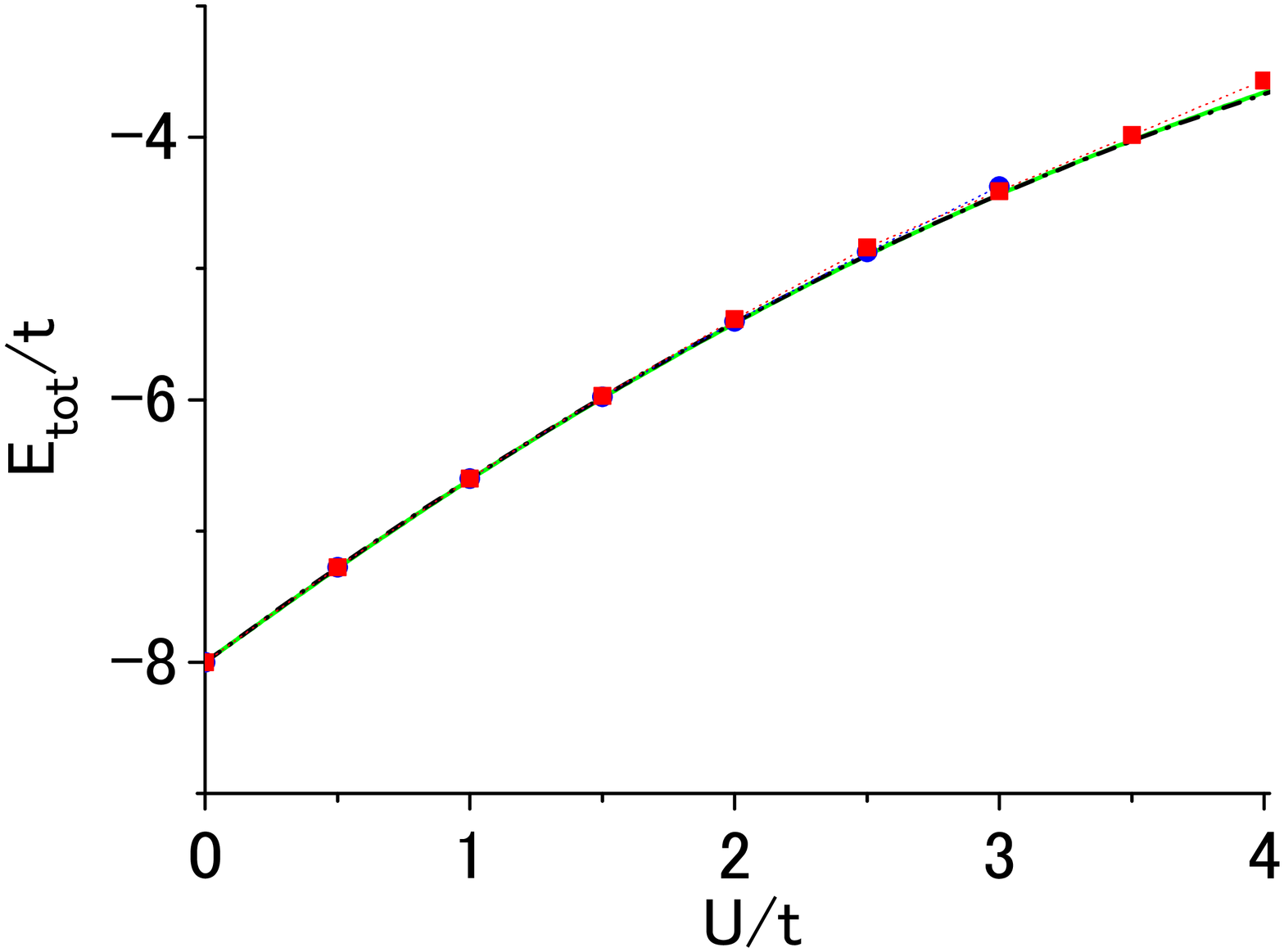}
}
\caption{Same as Fig. \ref{hubMF} but for the ground state energy $E_{\rm tot}$.}
\label{hubE}
\end{figure}

\begin{figure}
\resizebox{0.5\textwidth}{!}{%
\includegraphics{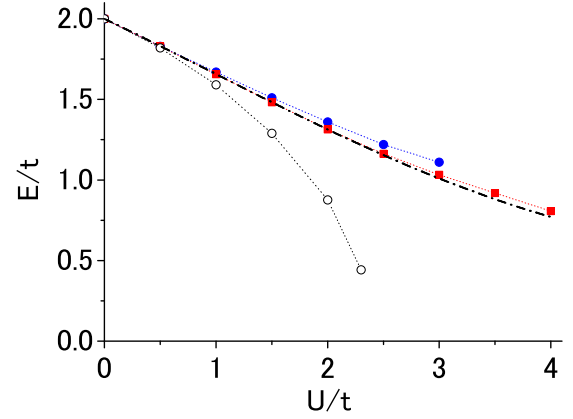}
}
\caption{Excitation energy of the first excited state calculated in SCRPA (circles) and ESRPA (squares) as a function of $U/t$ for the 
six-site Hubbard model with half-filling. The exact values are shown with the do-dashed line.The open circles depict the results in RPA.}
\label{hubex1}
\end{figure}

\subsection{Hubbard model}

Finally
we consider the one-dimensional (1D) Hubbard model with periodic boundary conditions.
In momentum space the Hamiltonian is given by
\begin{eqnarray}
H&=&\sum_{{\bm k},\sigma}\epsilon_{k}a^+_{{\bm k},\sigma}a_{{\bm k},\sigma}
\nonumber \\
&+&\frac{U}{2N}\sum_{{\bm k},{\bm p},{\bm q},\sigma}
a^+_{{\bm k},\sigma}a_{{\bm k}+{\bm q},\sigma}a^+_{{\bm p},-\sigma}a_{{\bm p}-{\bm q},-\sigma},
\end{eqnarray}
where 
$U$ is the on-site Coulomb matrix element, $\sigma$ spin projection and 
the single-particle energies are given by $\epsilon_k=-2t\sum_{d=1}^D\cos(k_d)$
with the nearest-neighbor hopping potential $t$.
We consider the case of six sites at half filling.
In the first Brillouin zone $-\pi\le k <\pi$ there are the following wave numbers
\begin{eqnarray}
k_1&=&0,~~~k_2=\frac{\pi}{3},~~~k_3=-\frac{\pi}{3},
\nonumber \\
k_4&=&\frac{2\pi}{3},~~~k_5=-\frac{2\pi}{3}.~~~k_6=-\pi.
\end{eqnarray}
The single-particle energies are $\epsilon_1=-2t$, $\epsilon_2=\epsilon_3=-t$, 
$\epsilon_4=\epsilon_5=t$ and $\epsilon_6=2t$. The ground state in TDDM is obtained using the adiabatic method starting from the HF
ground state where the six lowest-energy single-particle states are completely occupied: $T$ used here is $5\times 2\pi/t$.
The mean-field energy in ESRPA is calculated from the occupation probabilities given by Eq. (\ref{ninSTDDM})  
and the correlation energy in ESRPA given by Eq. (\ref{Eint3}): 
The first term on the right-hand side of Eq. (\ref{Eint4}) vanishes due to $p-h$ symmetry in the case of half-filling considered here.
In SCRPA all $p-h$ and $h-p$ amplitudes are taken and the factor 1/2 in Eq. (\ref{nh2}) is kept.  That is, we considered the following RPA excitation operator 

\begin{equation}
Q^+_{\nu} = \sum_{ph}[X^{\nu}_{ph}a^+_{p}a_{h} - Y^{\nu}_{ph}a^+_{h}a_{p}]
\end{equation}
where $p, h = ({\bm p},{\bm h}, \sigma )$ includes momenta and spin indices. Of course, in the end the SCRPA matrix will turn out to be block-diagonal in transferred momenta and charge and spin quantum numbers. However, in the set up of the SCRPA matrix, in the construction of the two body correlation functions all possible contributions are kept. Therefore, there is indirect coupling between all channels. This is different from \cite{Jemai} where the channels have been decoupled.

The matrix elements $C_{p_1p_2p_3p_4}$ and $C_{h_1h_2h_3h_4}$ are calculated using Eqs. (\ref{C4p}) and (\ref{C4h}) in SCRPA.
In the ESRPA calculations we take only the 2p-2h and 2h-2p
components of ${\mathcal X}^\mu_{\alpha\beta\alpha'\beta'}$ to facilitate the numerics. Since the three-body correlation matrix is an approximate one,
the stationary condition for the three-body correlation matrix is not completely fulfilled, which 
makes the Hamiltonian matrix of Eq. (\ref{SCS}) non-hermitian, especially in the case of the Hubbard model which has more general two-body interaction than
the Lipkin model and pairing models.

The mean-field energy $E_{\rm MF}$, which is given by 
\begin{eqnarray}
E_{\rm MF}=\sum_{{\bm k},\sigma} \epsilon_{\bm k}n_{{\bm k},\sigma}+\frac{U}{2N}\sum_{{\bm k},{\bm p},\sigma}n_{{\bm k},\sigma}n_{{\bm p},-\sigma},
\end{eqnarray}
the correlation energy $E_{\rm 2bcor}$ and the ground-state energy $E_{\rm tot}$ 
calculated in ESRPA (squares) are shown in Figs. \ref{hubMF}--\ref{hubE}
as a function of $U/t$. The results in SCRPA are also given with the circles up tp $U/t=3$ 
but cannot be distinguished from those in ESRPA. Beyond $U/t\approx 3$, SCRPA cannot give meaningful solutions because of numerical instabilities as is the case for the Lipkin model.
The TDDM results and the exact values are shown
with the solid and dot-dashed lines, respectively. The results in ESRPA agree well with those in TDDM.
The excitation energy of the first excited state is shown in Fig. \ref{hubex1} as a function of $U/t$. The results in ESRPA (squares) show good agreement with
the exact values (dot-dashed line). The SCRPA results (blue circles) are reasonable and avoid the instability of RPA (open circles). The SCRPA results in Fig. \ref{hubex1} are, however, less good than the ones in \cite{Jemai}. For the second excited state (not shown) the situation becomes even worse. This fact needs some discussion. The reason for the present SCRPA results for the excitation energies apparently is due to the implicit cross channel couplings meaning that in the block matrix belonging to, e.g., a certain momentum transfer $q$, implicitly via the non-linear terms other momentum transfers also can enter. In \cite{Jemai}, we discarded those 'intruder' channels for the following reason: since SCRPA does not strictly satisfy the killing condition (\ref{killing}), the various RPA operators $Q^+_{\nu}$ are not independent of one another. In \cite{Jemai2}, it was shown that this violation of independence is very weak. Apparently it is, however, still strong enough to perturb the equilibrium of the screening terms in the case of excitation energies. For correlation and ground state energies, the problem seems to be much less severe. It may, thus, be better to discard the implicit channel coupling for the excitation energies in SCRPA which is an approximation to ESRPA or STDDM* (STDDM*-b). The channel couplings can be restored if the two body amplitudes are taken care of as is seen in Fig.\ref{hubex1} what, however, renders the problem much harder to be solved.

\subsection{Summary of applications}
We applied TDDM, STDDM-b and STDDM*-b, ESRPA, SCRPA, and C-RPA to three exactly solvable models. We found that the ground-state properties obtained from the excites states in ESRPA 
agree well with those in TDDM and the exact values. This indicates that the TDDM equations (Eqs. (\ref{n}) and (\ref{N3C2})) build the ground state which is consistent with
excited states. We also found that the results of SCRPA (and C-RPA) agree with those in ESRPA except for strongly interacting regions where the systems enter a new phase as, e.g., superfluidity in the case of pairing or anti-ferromagnetism in the case of the Hubbard model.  
\noindent
\section{Conclusions}

In this work, we give a coherent outline of the BBGKY hierarchy or the time-dependent density matrix (TDDM) approach decoupled at the 3-body level in approximating the 3-body correlation function by a quadratic form of the 2-body correlation functions. The coupled equations for the 1-body density matrix and the 2-body correlation functions are then linearised around the equilibrium leading to eigenvalue equations coined STDDM-b and STDDM*-b which couple 1-body and 2-body amplitudes. The central part of the work is to show that the 1-body sector of the STDDM-b and STDDM*-b equations contains extended RPA equations which, contrary to standard RPA, are built on a correlated ground state. These extended RPA equations existed independently in the past and were called  Self-Consistent RPA (SCRPA) because the RPA matrix, due to the ground state correlations, depends on the 2-body correlation functions (screening terms) and, thus, a self-consistent cycle is needed for the solution. However, a second option is to take the 2-body correlation function (and the correlated occupation numbers) from an independent TDDM calculation for the ground state. This option has been coined correlated RPA (C-RPA). It was shown that the results in model cases are in both cases, as can be suspected, of similar quality. It was also shown in the present work that in SCRPA (C-RPA) equations  a very important part of the correlations contained in the full TDDM and STDDM-b (STDDM*-b) approaches is already incorporated. For instance, SCRPA (C-RPA) equations are, like TDDM non-linear in the 2-body correlations. This remark is very important from the practical point of view, since the dimension of the SCRPA (C-RPA) matrices is much reduced with respect to STDDM-b and STDDM*-b where the 2-body sector is included. The results for model cases show that at least for values of the coupling constants which are below or equal to the critical value where in mean field a phase transition occurs (exemple: BCS-pairing instability), the results from SCRPA (C-RPA) are practically of the same quality as the ones where the 2-body sector is included. We also could show that SCRPA fullfills all the desirable properties of standard RPA as there are: fullfilment of f-sum-rule, Goldstone (zero mode) theorem, conservation laws, and gauge invariance. It should be pointed out that those properties are usually very difficult to keep satisfied in beyond HF-RPA approaches with numerically manageable theories. For example the Kadanoff-Baym $\Phi$-derivable functional approach \cite{K-Baym} would face serious difficulties when applied to the models treated in this paper. SCRPA and C-RPA can also be applied in cases with a broken symmetry in allowing for symmetry broken mean field solutions. However, much less experience has accumulated in this regime. Only one work exists where the appearance of the Goldstone (zero) mode has been explicitly shown ~\cite{Lipkin3}. The results though good, suffer from the fact that the transition from the case with good symmetry to the one with broken symmetry is discontinuous, simulating a first order phase transition where there should not be any. A similar difficulty popped up with the Coupled Cluster Theory when applied to the pairing Hamiltonian ~\cite{pairing-cct1}. Very recently this difficulty of CCT has been circumvented in interpolating between the two regimes \cite{pairing-cct2}. Something similar is eventually also possible with TDDM and SCRPA. In any case the difficulty of artificial first order phase transition, see also \cite{NPA512}, probably arises from the fact that SCRPA is a truncated form of the more complete STDDM-b (STDDM*-b) theories. However, as mentioned, for practical (numerical) reasons, one would like to stay at the 1-body sector. For reasons of selfcontainedness, we repeated in this work some formalism already published elsewhere.\\
Other features to be pointed out concerning the present theory are that TDDM yields fully antisymmetric 2-body correlation functions and that they are number and energy conserving. They democratically couple particle-hole and particle-particle (hole-hole) channels and, thus, have some similarity with resummation of parquet diagrams. The quadratic form in the 2-body correlation functions is rather analogous, but one level higher, to the quadratic dependence of HF theory on the s.p. density matrix. Indeed the SCRPA  matrix may be viewed as the mean field Hamiltonien of density fluctuations. We also pointed out that SCRPA also exists in the particle-particle (hole-hole) channel and, then, the SCppRPA matrix can be interpreted as the mean field of, e.g., a two fermion bound state in an environment of those bound states. Our extended RPA equations are of the Schroedinger type and thus amenable to numerical solution. This is radically different from the usual many time Green's function formalism employed in several branches of physics.\\
It may be interesting to transform our TDDM equations with the non-linear decoupling of the 3-body correlations into classical transport equations. This shall be work for the future.

\acknowledgments

We are very greatful to D. Delion, J. Dukelsky, M. Jemai, and A. Storozhenko for previous collaborations on the subject of SCRPA. The present new developments are based on those earlier works.

\vspace{2cm}

\appendix
\section{Quadratic form of $C_3$}
\subsection{Green's function}

In the first part of this section, we give a sketchy derivation using many body Green's functions how $C_3$ can be expressed as a quadratic form of $C_2$'s. In the second part a much more formal derivation will be given using identites of many particle density matrices.\\

The one line reducible part of the $2p-1h (2h-1p)$ Green's function can be written as \cite{RS}, App. F, see also Fig.1

\begin{equation}
G^{corr,t-t'}_{\alpha \beta \gamma' \alpha' \beta' \gamma}  = \int dt_{\delta}\int dt_{\delta'}G^{t-t_{\delta}}_{\alpha \beta \gamma' j_{\delta}}G^{t_{\delta}-t_{\delta'}}_{\delta \delta'}
G^{t_{\delta'}-t'}_{j_{\delta'}\gamma \beta' \alpha'}
\label{G3corr}
\end{equation}

\noindent
with 

\begin{equation}
G^{t-t_1}_{\alpha \beta \gamma' j_{\delta_1}}=(-i)\langle 0|T(a^+_{\gamma'}a_{\alpha}a_{\beta})_tj^+_{\delta_1, t_1}|0\rangle_{\rm irr}
\label{G2j}
\end{equation}
and $  j^+_{\delta_1, t_1}=\frac{1}{2}\sum (a^+_{\gamma}a^+_{\delta}a_{\beta})_{t_1}\bar v_{\delta\gamma \beta \delta_1}$. The index 
'irr' stands for 'one line irreducible'. For convenience, we make for the single particle Green's function the quasi-particle approximation 

\begin{equation}
G^{t-t'_{\delta}}_{\delta \delta'}= \delta_{\delta \delta'}\{\bar n_{\delta}\Theta(t-t') - n_{ \delta}\Theta(t'-t)\}e^{-i\epsilon_{\delta}(t-t')}
\label{Gqp}
\end{equation}
where the $n_{\delta}$ are the correlated quasi-particle occupation numbers and $\epsilon_{\delta}$ is the energy of the quasi-particle pole of the s.p. Green's function, see \cite{RS}. Let us now insert this quasi-particle expression into (\ref{G3corr}) and take the equal time limit, so that the 3-body correlation function $C_3$ appears. We will see that the quadratic form in $C_2$'s is obtained. \\
This derivation is kept very qualitative, just to give the reader an impression how from the very natural expression for (\ref{G3corr}) for the one line reducible part of the $2p-1h (2h-1p)$ Green's function our quadratic form for $C_3$ in terms of $C_2$'s can appear. Let us now switch to a mathematically very transparent derivation involving identies of density matrices.

\subsection{Identity}
The quadratic form of the three-body correlation matrix is also obtained from the identity between three-body and four-body density matrices:
\begin{eqnarray}
\rho_{\alpha\beta\gamma\alpha'\beta'\gamma'}=\frac{1}{N-3}\sum_\lambda\rho_{\alpha\beta\gamma\lambda\alpha'\beta'\gamma'\lambda}.
\end{eqnarray}
The above identity is written in terms of correlation matrices as
\begin{eqnarray}
C_{\alpha\beta\gamma\alpha'\beta'\gamma'}&=&\frac{1}{3}\sum_\lambda(n_{\alpha\lambda}C_{\lambda\beta\gamma\alpha'\beta'\gamma'}
+n_{\beta\lambda}C_{\alpha\lambda\gamma\alpha'\beta'\gamma'}
\nonumber \\
&+&n_{\gamma\lambda}C_{\alpha\beta\lambda\alpha'\beta'\gamma'}
+n_{\lambda\alpha'}C_{\alpha\beta\gamma\lambda\beta'\gamma'}
\nonumber \\
&+&n_{\lambda\beta'}C_{\alpha\beta\gamma\alpha'\lambda\gamma'}
+n_{\lambda\gamma'}C_{\alpha\beta\gamma\alpha'\beta'\lambda}
\nonumber \\
&-&C_{\alpha\beta\alpha'\lambda}C_{\gamma\lambda\beta'\gamma'}
-C_{\alpha\beta\gamma'\lambda}C_{\gamma\lambda\alpha'\beta'}
\nonumber \\
&-&C_{\alpha\beta\lambda\beta'}C_{\gamma\lambda\alpha'\gamma'}
-C_{\alpha\gamma\alpha'\lambda}C_{\lambda\beta\beta'\gamma'}
\nonumber \\
&-&C_{\alpha\gamma\beta'\lambda}C_{\beta\lambda\alpha'\gamma'}
-C_{\alpha\gamma\lambda\gamma'}C_{\beta\lambda\alpha'\beta'}
\nonumber \\
&-&C_{\alpha\lambda\alpha'\beta'}C_{\beta\gamma\gamma'\lambda}
-C_{\alpha\lambda\alpha'\gamma'}C_{\beta\gamma\lambda\beta'}
\nonumber \\
&-&C_{\alpha\lambda\beta'\gamma'}C_{\beta\gamma\alpha'\lambda}-C_{\alpha\beta\gamma\lambda\alpha'\beta'\gamma'\lambda}),
\label{3body}
\end{eqnarray}
where $C_{\alpha\beta\gamma\lambda\alpha'\beta'\gamma'\lambda}$ is a four-body correlation matrix.
Under the assumptions that $n_{\alpha\alpha'}=\delta_{\alpha\alpha'}n_\alpha$ and $C_{\alpha\beta\gamma\lambda\alpha'\beta'\gamma'\lambda}=0$,
the above relation is given as
\begin{eqnarray}
C_{\alpha\beta\gamma\alpha'\beta'\gamma'}&=&\frac{1}{3-n_{\alpha}-n_{\beta}-n_{\gamma}-n_{\alpha'}-n_{\beta'}-n_{\gamma'}}
\nonumber \\
&\times&\sum_\lambda(
-C_{\alpha\beta\alpha'\lambda}C_{\gamma\lambda\beta'\gamma'}
-C_{\alpha\beta\gamma'\lambda}C_{\gamma\lambda\alpha'\beta'}
\nonumber \\
&-&C_{\alpha\beta\lambda\beta'}C_{\gamma\lambda\alpha'\gamma'}
-C_{\alpha\gamma\alpha'\lambda}C_{\lambda\beta\beta'\gamma'}
\nonumber \\
&-&C_{\alpha\gamma\beta'\lambda}C_{\beta\lambda\alpha'\gamma'}
-C_{\alpha\gamma\lambda\gamma'}C_{\beta\lambda\alpha'\beta'}
\nonumber \\
&-&C_{\alpha\lambda\alpha'\beta'}C_{\beta\gamma\gamma'\lambda}
-C_{\alpha\lambda\alpha'\gamma'}C_{\beta\gamma\lambda\beta'}
\nonumber \\
&-&C_{\alpha\lambda\beta'\gamma'}C_{\beta\gamma\alpha'\lambda}).
\label{3body1}
\end{eqnarray}
For $C_{p_1h_1h_2,p_2h_3h_4}$ the denominator of Eq. (\ref{3body1}) is $-1$ and Eq. (\ref{3body2}) is obtained.

\section{Matrices in STDDM}
The matrices $a=$, $b$, $c$, $d$ and $\Delta d$
in Eq. (\ref{STDDM0}) are given below.
\begin{eqnarray}
a(\alpha\alpha':\lambda\lambda')&=&(\epsilon_{\alpha}-\epsilon_{\alpha'})\delta_{\alpha\lambda}\delta_{\alpha'\lambda'}
\nonumber \\
&+&\sum_{\beta}(
\bar{v}_{\alpha\lambda'\beta\lambda}n_{\beta\alpha'}-
\bar{v}_{\beta\lambda'\alpha'\lambda}n_{\alpha\beta}),
\label{matrixa}
\end{eqnarray}
\begin{eqnarray}
b(\alpha\alpha':\lambda_1\lambda_2\lambda_1'\lambda_2')&=&\frac{1}{2}(
\bar{v}_{\alpha\lambda_2'\lambda_1\lambda_2}\delta_{\alpha'\lambda_1'}
-\bar{v}_{\lambda_1'\lambda_2'\alpha'\lambda_2}\delta_{\alpha\lambda_1}),
\nonumber \\
\label{matrixb}
\end{eqnarray}
\begin{eqnarray}
&c(&\alpha_1\alpha_2\alpha_1'\alpha_2':\lambda\lambda')
=
-\delta_{\alpha_1\lambda}\{\sum_{\beta\gamma\delta}[(\delta_{\alpha_2\beta}-n_{\alpha_2\beta})
n_{\gamma\alpha_1'}n_{\delta\alpha_2'}
\nonumber \\
&+&n_{\alpha_2\beta}(\delta_{\gamma\alpha_1'}-n_{\gamma\alpha_1'})(\delta_{\delta\alpha_2'}-n_{\delta\alpha_2'})]
\bar{v}_{\lambda'\beta\gamma\delta}
\nonumber \\
&+&\sum_{\beta\gamma}[\frac{1}{2}\bar{v}_{\lambda'\alpha_2\beta\gamma} C_{\beta\gamma\alpha_1'\alpha_2'}
+\bar{v}_{\lambda'\beta\alpha_1'\gamma} C_{\alpha_2\gamma\alpha_2'\beta}
\nonumber \\
&-&\bar{v}_{\lambda'\beta\alpha_2'\gamma} C_{\alpha_2\gamma\alpha_1'\beta}]\}
\nonumber \\
&+&\delta_{\alpha_2\lambda}\{\sum_{\beta\gamma\delta}[(\delta_{\alpha_1\beta}-n_{\alpha_1\beta})
n_{\gamma\alpha_1'}n_{\delta\alpha_2'}
\nonumber \\
&+&n_{\alpha_1\beta}(\delta_{\gamma\alpha_1'}-n_{\gamma\alpha_1'})(\delta_{\delta\alpha_2'}-n_{\delta\alpha_2'})]
\bar{v}_{\lambda'\beta\gamma\delta}
\nonumber \\
&+&\sum_{\beta\gamma}[\frac{1}{2}\bar{v}_{\lambda'\alpha_1\beta\gamma} C_{\beta\gamma\alpha_1'\alpha_2'}
+\bar{v}_{\lambda'\beta\alpha_1'\gamma} C_{\alpha_1\gamma\alpha_2'\beta}
\nonumber \\
&-&\bar{v}_{\lambda'\beta\alpha_2'\gamma} C_{\alpha_1\gamma\alpha_1'\beta}]\}
\nonumber \\
&+&\delta_{\alpha_1'\lambda'}\{\sum_{\beta\gamma\delta}[(\delta_{\delta\alpha_2'}-n_{\delta\alpha_2'})
n_{\alpha_1\beta}n_{\alpha_2\gamma}
\nonumber \\
&+&n_{\delta\alpha_2'}
(\delta_{\alpha_1\beta}-n_{\alpha_1\beta})(\delta_{\alpha_2\gamma}-n_{\alpha_2\gamma})]
\bar{v}_{\beta\gamma|v|\lambda\delta}
\nonumber \\
&+&\sum_{\beta\gamma}[\frac{1}{2}\bar{v}_{\beta\gamma\lambda\alpha_2'} C_{\alpha_1\alpha_2\beta\gamma}
+\bar{v}_{\alpha_1\beta\lambda\gamma} C_{\alpha_2\gamma\alpha_2'\beta}
\nonumber \\
&-&\bar{v}_{\alpha_2\beta\lambda\gamma} C_{\alpha_1\gamma\alpha_2'\beta}]\}
\nonumber \\
&-&\delta_{\alpha_2'\lambda'}\{\sum_{\beta\gamma\delta}[(\delta_{\delta\alpha_1'}-n_{\delta\alpha_1'})
n_{\alpha_1\beta}n_{\alpha_2\gamma}
\nonumber \\
&+&n_{\delta\alpha_1'}
(\delta_{\alpha_1\beta}-n_{\alpha_1\beta})(\delta_{\alpha_2\gamma}-n_{\alpha_2\gamma})]
\bar{v}_{\beta\gamma\lambda\delta}
\nonumber \\
&+&\sum_{\beta\gamma}[\frac{1}{2}\bar{v}_{\beta\gamma\lambda\alpha_1'} C_{\alpha_1\alpha_2\beta\gamma}
+\bar{v}_{\alpha_1\beta\lambda\gamma} C_{\alpha_2\gamma\alpha_1'\beta}
\nonumber \\
&-&\bar{v}_{\alpha_2\beta\lambda\gamma} C_{\alpha_1\gamma\alpha_1'\beta}]\}
\nonumber \\
&+&\sum_{\beta}[\bar{v}_{\alpha_1\lambda'\beta\lambda} C_{\beta\alpha_2\alpha_1'\alpha_2'}
-\bar{v}_{\alpha_2\lambda'\beta\lambda} C_{\beta\alpha_1\alpha_1'\alpha_2'}
\nonumber \\
&-&\bar{v}_{\beta\lambda'\alpha_2'\lambda} C_{\alpha_1\alpha_2\alpha_1'\beta}
+\bar{v}_{\beta\lambda'\alpha_1'\lambda} C_{\alpha_1\alpha_2\alpha_2'\beta}],
\label{matrxb}
\end{eqnarray}
\begin{eqnarray}
&d(&\alpha_1\alpha_2\alpha_1'\alpha_2':\lambda_1\lambda_2\lambda_1'\lambda_2')=(\epsilon_{\alpha_1}+\epsilon_{\alpha_2}-\epsilon_{\alpha_1'}-\epsilon_{\alpha_2'})
\nonumber \\
&\times&\delta_{\alpha_1\lambda_1}\delta_{\alpha_2\lambda_2}
\delta_{\alpha_1'\lambda_1'}\delta_{\alpha_2'\lambda_2'}
\nonumber \\
&+&\frac{1}{2}\delta_{\alpha_1'\lambda_1'}\delta_{\alpha_2'\lambda_2'}
\sum_{\beta\gamma}(\delta_{\alpha_1\beta}\delta_{\alpha_2\gamma}
-\delta_{\alpha_2\gamma}n_{\alpha_1\beta}
-\delta_{\alpha_1\beta}n_{\alpha_2\gamma})
\nonumber \\
&\times&\bar{v}_{\beta\gamma\lambda_1\lambda_2}
\nonumber \\
&-&\frac{1}{2}\delta_{\alpha_1\lambda_1}\delta_{\alpha_2\lambda_2}
\sum_{\beta\gamma}(\delta_{\alpha_1'\beta}\delta_{\alpha_2'\gamma}
-\delta_{\alpha_2'\gamma}n_{\beta\alpha_1'}
-\delta_{\alpha_1'\beta}n_{\gamma\alpha_2'})
\nonumber \\
&\times&\bar{v}_{\lambda_1'\lambda_2'\beta\gamma}
\nonumber \\
&+&\delta_{\alpha_2\lambda_2}\delta_{\alpha_2'\lambda_2'}
\sum_{\beta}(\bar{v}_{\alpha_1\lambda_1'\beta\lambda_1}n_{\beta\alpha_1'}
-\bar{v}_{\beta\lambda_1'\alpha_1'\lambda_1}n_{\alpha_1\beta})
\nonumber \\
&+&\delta_{\alpha_2\lambda_2}\delta_{\alpha_1'\lambda_1'}
\sum_{\beta}(\bar{v}_{\alpha_1\lambda_2'\beta\lambda_1}n_{\beta\alpha_2'}
-\bar{v}_{\beta\lambda_2'\alpha_2'\lambda_1}n_{\alpha_1\beta})
\nonumber \\
&+&\delta_{\alpha_1\lambda_1}\delta_{\alpha_1'\lambda_1'}
\sum_{\beta}(\bar{v}_{\alpha_2\lambda_2'\beta\lambda_2}n_{\beta\alpha_2'}
-\bar{v}_{\beta\lambda_2'\alpha_2'\lambda_2}n_{\alpha_2\beta})
\nonumber \\
&+&\delta_{\alpha_1\lambda_1}\delta_{\alpha_2'\lambda_2'}
\sum_{\beta}(\bar{v}_{\alpha_2\lambda_1'\beta\lambda_2}n_{\beta\alpha_1'}
-\bar{v}_{\beta\lambda_1'\alpha_1'\lambda_2}n_{\alpha_2\beta}).
\label{matrixd}
\end{eqnarray}

We now give the expression for
$\Delta d$ which arises from the quadratic forms in $C_2$'s of the 3-body correlation functions. We use Eqs. (\ref{C_2C_2-h}) and (\ref{C_2C_2-p}) for the three-body
correlation matrix.
\begin{eqnarray}
&\Delta d(&\alpha\beta\alpha'\beta':\lambda_1\lambda_2\lambda_1'\lambda_2')=
-\frac{1}{2}\bar{v}_{\alpha(h)\lambda_1'(h)\lambda_1(p)\lambda_2(p)}
\nonumber \\
&\times& C_{\lambda_2'(h)\beta(h)\alpha'(p)\beta'(p)}
\nonumber \\
&+&\frac{1}{2}\bar{v}_{\alpha(p)\lambda_1'(p)\lambda_1(h)\lambda_2(h)}C_{\lambda_2'(p)\beta(p)\alpha'(h)\beta'(h)}
\nonumber \\
&-&\frac{1}{2}\delta_{\beta\lambda_2}\delta_{\alpha'\lambda_1'}\delta_{\beta'\lambda_2'}
\sum_{\lambda(h)\lambda'(p)\lambda''(p)}\bar{v}_{\alpha\lambda\lambda'\lambda''}C_{\lambda'\lambda''\lambda\lambda_1(h)}
\nonumber \\
&-&\frac{1}{2}\delta_{\beta\lambda_1}\delta_{\alpha'\lambda_1'}\delta_{\beta'\lambda_2'}
\sum_{\lambda(p)\lambda'(h)\lambda''(h)}\bar{v}_{\alpha\lambda\lambda'\lambda''}C_{\lambda'\lambda''\lambda\lambda_2(p)}
\nonumber \\
&+&\frac{1}{2}\bar{v}_{\beta(h)\lambda_1'(h)\lambda_1(p)\lambda_2(p)}C_{\lambda_2'(h)\alpha(h)\alpha'(p)\beta'(p)}
\nonumber \\
&-&\frac{1}{2}\bar{v}_{\beta(p)\lambda_1'(p)\lambda_1(h)\lambda_2(h)}C_{\lambda_2'(p)\alpha(p)\alpha'(h)\beta'(h)}
\nonumber \\
&+&\frac{1}{2}\delta_{\alpha\lambda_2}\delta_{\alpha'\lambda_1'}\delta_{\beta'\lambda_2'}
\sum_{\lambda(h)\lambda'(p)\lambda''(p)}\bar{v}_{\beta\lambda\lambda'\lambda''}C_{\lambda'\lambda''\lambda\lambda_1(h)}
\nonumber \\
&+&\frac{1}{2}\delta_{\alpha\lambda_1}\delta_{\alpha'\lambda_1'}\delta_{\beta'\lambda_2'}
\sum_{\lambda(p)\lambda'(h)\lambda''(h)}\bar{v}_{\beta\lambda\lambda'\lambda''}C_{\lambda'\lambda''\lambda\lambda_2(p)}
\nonumber \\
&+&\frac{1}{2}\bar{v}_{\lambda_1'(p)\lambda_2'(p)\alpha'(h)\lambda_2(h)}C_{\alpha(p)\beta(p)\beta'(h)\lambda_1(h)}
\nonumber \\
&-&\frac{1}{2}\bar{v}_{\lambda_1'(h)\lambda_2'(h)\alpha'(p)\lambda_2(p)}C_{\alpha(h)\beta(h)\beta'(p)\lambda_1(p)}
\nonumber \\
&+&\frac{1}{2}\delta_{\alpha\lambda_1}\delta_{\beta\lambda_2}\delta_{\beta'\lambda_1'}
\sum_{\lambda(p)\lambda'(p)\lambda''(h)}\bar{v}_{\lambda\lambda'\alpha'\lambda''}C_{\lambda_2'(h)\lambda''\lambda\lambda'}
\nonumber \\
&-&\frac{1}{2}\delta_{\alpha\lambda_1}\delta_{\beta\lambda_2}\delta_{\beta'\lambda_1'}
\sum_{\lambda(h)\lambda'(h)\lambda''(p)}\bar{v}_{\lambda\lambda'\alpha'\lambda''}C_{\lambda_2'(p)\lambda''\lambda\lambda'}
\nonumber \\
&-&\frac{1}{2}\bar{v}_{\lambda_1'(p)\lambda_2'(p)\beta'(h)\lambda_2(h)}C_{\alpha(p)\beta(p)\alpha'(h)\lambda_1(h)}
\nonumber \\
&+&\frac{1}{2}\bar{v}_{\lambda_1'(h)\lambda_2'(h)\beta'(p)\lambda_2(p)}C_{\alpha(h)\beta(h)\alpha'(p)\lambda_1(p)}
\nonumber \\
&-&\frac{1}{2}\delta_{\alpha\lambda_1}\delta_{\beta\lambda_2}\delta_{\alpha'\lambda_1'}
\sum_{\lambda(p)\lambda'(p)\lambda''(h)}\bar{v}_{\lambda\lambda'\beta'\lambda''}C_{\lambda_2'(h)\lambda''\lambda\lambda'}
\nonumber \\
&+&\frac{1}{2}\delta_{\alpha\lambda_1}\delta_{\beta\lambda_2}\delta_{\alpha'\lambda_1'}
\sum_{\lambda(h)\lambda'(h)\lambda''(p)}\bar{v}_{\lambda\lambda'\beta'\lambda''}C_{\lambda_2'(p)\lambda''\lambda\lambda'}
\nonumber \\
\end{eqnarray}
The terms with and without summation describe self-energy corrections and vertex corrections, respectively, and indices
$p$ ($h)$ mean that the corresponding single-particle state is a particle (hole) state.

\section{}

The matrix $E_{\alpha\beta\alpha'\beta'}$ is the product of the self-energy terms in Eq. (\ref{S-term}) and the first term on rhs in Eq. (\ref{D}) \\
\begin{eqnarray}
E_{\alpha\beta\alpha'\beta'}&=&-\frac{1}{4}\sum_{\lambda\lambda'\lambda''}[\bar{v}_{\alpha\lambda\lambda'\lambda''} C_{\lambda'\lambda''\beta'\lambda}
\nonumber \\
&\times&\delta_{\beta\alpha'}\left(L_{\alpha'\beta'}-L_{\alpha'\alpha}\right)
\nonumber \\
&-&\bar{v}_{\beta\lambda\lambda'\lambda''} C_{\lambda'\lambda''\beta'\lambda}
\delta_{\alpha\alpha'}\left(L_{\alpha'\beta'}-L_{\alpha'\beta}\right)
\nonumber \\
&+&\bar{v}_{\lambda\lambda'\alpha'\lambda''} C_{\beta\lambda''\lambda\lambda'}
\delta_{\alpha\beta'}\left(L_{\beta\alpha}-L_{\alpha'\alpha}\right)
\nonumber \\
&-&\bar{v}_{\lambda\lambda'\alpha'\lambda''} C_{\alpha\lambda''\lambda\lambda'}
\delta_{\beta\beta'}\left(L_{\alpha\beta}-L_{\alpha'\beta}\right)],
\label{E}
\end{eqnarray}
where $L_{\alpha \beta}=\frac{n_{\alpha}\bar n_{\beta}}{n_{\alpha} - n_{\beta}}$. 
The matrix $E_{\alpha\beta\alpha'\beta'}$  describes the self-energy corrections but has no contribution to
the 2p-2h and 2h-2p elements of $C_{\alpha\beta\alpha'\beta'}$.

The matrix $F_{\alpha\beta\alpha'\beta'}$ is obtained from the product of the vertex correction terms in Eq. (\ref{S-term}) and the first term on rhs in Eq. (\ref{D}).
\begin{eqnarray}
F_{\alpha\beta\alpha'\beta'}&=&\frac{1}{2}\sum_{\lambda\lambda'}
[(\bar{v}_{\alpha\lambda\beta'\lambda'}C_{\beta\lambda'\alpha'\lambda}
+\bar{v}_{\beta\lambda\alpha'\lambda'}C_{\alpha\lambda'\beta'\lambda})
\nonumber \\
&\times&\left(L_{\beta\beta'}-L_{\alpha'\alpha}\right)
\nonumber \\
&-&(\bar{v}_{\beta\lambda\beta'\lambda'}C_{\alpha\lambda'\alpha'\lambda}
+\bar{v}_{\alpha\lambda\alpha'\lambda'}C_{\beta\lambda'\beta'\lambda})
\nonumber \\
&\times&
\left(L_{\alpha\beta'}-L_{\alpha'\beta}\right)]
\nonumber \\
&-&\frac{1}{4}\sum_{\lambda\lambda'}(\bar{v}_{\alpha\beta\lambda\lambda'} C_{\lambda\lambda'\alpha'\beta'}
+\bar{v}_{\lambda\lambda'\alpha'\beta'} C_{\alpha\beta\lambda\lambda'})
\nonumber \\
&\times&
(L_{\beta\beta'}+L_{\alpha\beta'}-L_{\alpha'\alpha}-L_{\alpha'\beta}).
\label{F}
\end{eqnarray}
As shown below, the factor $1/2$ in front of the $p-h$ correlation terms in Eq. (\ref{gs}) is dropped due to the $p-h$ correlation terms of $F_{p_1p_2h_1h_2}$   in Eq. (\ref{F}) (the first sum)
in the case that $n_\alpha=0$ or 1.
The $p-p$ and $h-h$ correlations are also included in Eq. (\ref{F}) (the last sum).

The matrix $G_{\alpha\beta\alpha'\beta'}$ is obtained from the product of the terms with $C_{\alpha\beta\alpha'\beta'}$ in Eq. (\ref{S-term}) and the second term on rhs in Eq. (\ref{D}).
\begin{eqnarray}
G_{\alpha\beta\alpha'\beta'}&=&-\frac{1}{4}\sum_{\lambda\lambda'\lambda''\gamma}[\bar{v}_{\alpha\lambda\lambda'\lambda''} C_{\lambda'\lambda''\gamma\lambda}
U_{\alpha'\gamma}C_{\gamma\beta\alpha'\beta'}
\nonumber \\
&-&\bar{v}_{\beta\lambda\lambda'\lambda''} C_{\lambda'\lambda''\gamma\lambda}
U_{\alpha'\gamma}C_{\gamma\alpha\alpha'\beta'}
\nonumber \\
&+&\bar{v}_{\lambda\lambda'\alpha'\lambda''} C_{\gamma\lambda''\lambda\lambda'}
U_{\gamma\alpha}C_{\alpha\beta\gamma\beta'}
\nonumber \\
&-&\bar{v}_{\lambda\lambda'\alpha'\lambda''} C_{\gamma\lambda''\lambda\lambda'}
U_{\gamma\beta}C_{\beta\alpha\gamma\beta'}]
\nonumber \\
&+&\frac{1}{4}\sum_{\lambda\lambda'\lambda''\gamma}[
\bar{v}_{\lambda'\lambda''\beta'\lambda} C_{\gamma\lambda\lambda'\lambda''}
U_{\beta\gamma}C_{\alpha\beta\alpha'\gamma}
\nonumber \\
&+&
\bar{v}_{\beta\lambda''\lambda\lambda'} C_{\lambda\lambda'\gamma\lambda''}
U_{\gamma\beta'}C_{\alpha\gamma\alpha'\beta'}
\nonumber \\
&-&
\bar{v}_{\lambda'\lambda''\beta'\lambda} C_{\gamma\lambda\lambda'\lambda''}
U_{\alpha\gamma}C_{\beta\alpha\alpha'\gamma}
\nonumber \\
&-&
\bar{v}_{\alpha\lambda''\lambda\lambda'} C_{\lambda\lambda'\gamma\lambda''}
U_{\gamma\beta'}C_{\beta\gamma\alpha'\beta'}]
\nonumber \\
&+&\frac{1}{2}\sum_{\lambda\lambda'\lambda''\gamma}[\bar{v}_{\alpha\lambda\lambda'\lambda''} C_{\gamma\lambda''\alpha'\lambda}
U_{\gamma\lambda'}C_{\lambda'\beta\gamma\beta'}
\nonumber \\
&-&\bar{v}_{\beta\lambda\lambda'\lambda''} C_{\gamma\lambda''\alpha'\lambda}
U_{\gamma\lambda'}C_{\lambda'\alpha\gamma\beta'}
\nonumber \\
&+&\bar{v}_{\lambda\lambda'\alpha'\lambda''} C_{\alpha\lambda''\gamma\lambda'}
U_{\lambda\gamma}C_{\gamma\beta\lambda\beta'}
\nonumber \\
&-&\bar{v}_{\lambda\lambda'\alpha'\lambda''} C_{\beta\lambda''\gamma\lambda'}
U_{\lambda\gamma}C_{\gamma\alpha\lambda\beta'}]
\nonumber \\
&-&\frac{1}{4}\sum_{\lambda\lambda'\lambda''\gamma}[\bar{v}_{\alpha\lambda''\lambda\lambda'} C_{\lambda\lambda'\alpha'\gamma}
U_{\lambda''\gamma}C_{\gamma\beta\lambda''\beta'}
\nonumber \\
&-&\bar{v}_{\beta\lambda''\lambda\lambda'} C_{\lambda\lambda'\alpha'\gamma}
U_{\lambda''\gamma}C_{\gamma\alpha\lambda''\beta'}
\nonumber \\
&+&\bar{v}_{\lambda\lambda'\alpha'\lambda''} C_{\alpha\gamma\lambda\lambda'}
U_{\gamma\lambda''}C_{\lambda''\beta\gamma\beta'}
\nonumber \\
&-&\bar{v}_{\lambda\lambda'\alpha'\lambda''} C_{\beta\gamma\lambda\lambda'}
U_{\gamma\lambda''}C_{\lambda''\alpha\gamma\beta'}
\nonumber \\
&-&\frac{1}{2}\sum_{\lambda\lambda'\lambda''\gamma}[\bar{v}_{\lambda\lambda'\beta'\lambda''} C_{\beta\lambda''\gamma\lambda'}
U_{\gamma\lambda}C_{\alpha\gamma\alpha'\lambda}
\nonumber \\
&-&\bar{v}_{\lambda\lambda'\beta'\lambda''} C_{\alpha\lambda''\gamma\lambda'}
U_{\gamma\lambda}C_{\beta\gamma\alpha'\lambda}
\nonumber \\
&+&\bar{v}_{\beta\lambda\lambda'\lambda''} C_{\gamma\lambda''\beta'\lambda}
U_{\lambda'\gamma}C_{\alpha\lambda'\alpha'\gamma}
\nonumber \\
&-&\bar{v}_{\alpha\lambda\lambda'\lambda''} C_{\gamma\lambda''\beta'\lambda}
U_{\lambda'\gamma}C_{\beta\lambda'\alpha'\gamma}]
\nonumber \\
&+&\frac{1}{4}\sum_{\lambda\lambda'\lambda''\gamma}[\bar{v}_{\lambda\beta\lambda'\lambda''} C_{\lambda'\lambda''\gamma\beta'}
U_{\gamma\lambda}C_{\alpha\gamma\alpha'\lambda}
\nonumber \\
&-&\bar{v}_{\lambda\alpha\lambda'\lambda''} C_{\lambda'\lambda''\gamma\beta'}
U_{\gamma\lambda}C_{\beta\gamma\alpha'\lambda}
\nonumber \\
&+&\bar{v}_{\lambda\lambda'\lambda''\beta'} C_{\gamma\beta\lambda\lambda'}
U_{\lambda''\gamma}C_{\alpha\lambda''\alpha'\gamma}
\nonumber \\
&-&\bar{v}_{\lambda\lambda'\lambda''\beta'} C_{\gamma\alpha\lambda\lambda'}
U_{\lambda''\gamma}C_{\beta\lambda''\alpha'\gamma}
],
\label{G}
\end{eqnarray}
where $U_{\alpha\beta}=\frac{1}{n_{\alpha}-n_{\beta}}$.
Here, we used the stationary condition for $n_\alpha$ ((Eq. (\ref{n})) in the second sum, which is given by 
\begin{eqnarray}
\sum_{\lambda_1\lambda_2\lambda_3}
[\bar{v}_{\alpha\lambda_1\lambda_2\lambda_3} C_{\lambda_2\lambda_3\alpha'\lambda_1}
&-&C_{\alpha\lambda_1\lambda_2\lambda_3}\bar{v}_{\lambda_2\lambda_3\alpha'\lambda_1}]=0.
\nonumber \\
\label{n1}
\end{eqnarray}
The terms in Eq. (\ref{G}) describe the contributions of the three-body correlation matrix $C_{\alpha\beta\gamma\alpha'\beta'\gamma}$.
Comparing Eq. (\ref{G}) with Eq. (\ref{3body2}), we notice that there is a factor 2 difference and that the term with $C_{\lambda_2\gamma\alpha'\beta'}C_{\lambda_3\beta\lambda_1\gamma}$
is missing in Eq. (\ref{G}). So, Eq. (\ref{gs}) is very similar to Eq. (\ref{N3C2}) 
but there are differences in the three-body terms.


\begin{thebibliography}{99}

\bibitem{Perdew}
J. P. Perdew, M. Levy, Phys. Rev. Lett. {\bf 51},1884 (1983).
\bibitem{BHF}
W. Zuo, I. Bombaci, U. Lombardo, Phys. Rev. C {\bf 60}, 024605 (1999).
\bibitem{Gutz}
P. Fulde, {\it Electron correlations in molecules and solids}, Springer Series in Solid state Sciences, 100. Springer, Berlin 1991.
\bibitem{CCT}
P. J. Knowles, C. Hampel, H.-J. Werner, J. Chem. Phys. {\bf 99}, 5219 (1993).
\bibitem{Bishop}
R. F. Bishop, Theor. Chim. Acta, {\bf 80}, 95 (1991).
\bibitem{QMC}
M. P. Nightingale, C. J. Umrigar, (Eds.) {\it Quantum Monte Carlo Methods in Physics and Chemistry}, (Springer, Berlin, 1999).
\bibitem{Markus}
M. Holzmann, B. Bernu, C. Pierleoni, J. Mc Minis, D. M. Ceperly, V. Olevano, L. Delle Site, Phys. Rev. Lett. {\bf 107},110402 (2011).
\bibitem{DMRG}
I. Peschel, X. Wang, M. Kaulke, K. Hallberg (Eds.), {\it Density-Matrix Renormalisation, A new Numerical Method in Physics}, (Springer, Berlin, 1999).
\bibitem{Schollwoeck}
U. Schollwoeck, Ann. Physics {\bf 326}, 96 (2011) {\it and} Rev. Mod. Phys. {\bf 77}, 259 (2005).
\bibitem{Clark}
J. W. Clark, P. Westhaus, Phys. Rev. {\bf 141}, 833 91966).
\bibitem{NPA628}
J. Dukelsky, G. Roepke, P. Schuck, Nucl. Phys. A {\bf 628}, 17 (1998).
\bibitem{NPA512}
J. Dukelsky, P. Schuck, Nucl. Phys. A {\bf 512}, 466 (1990).
\bibitem{Delion}
D. S. Delion, P. Schuck, J. Dukelsky, Phys. Rev. C {\bf 71}, 064305 (2005).
\bibitem{Hirsch}
J. G. Hirsch, A. Mariano, J. Dukelsky, P. Schuck, Ann. Physics {\bf 296}, 187 (2002).
\bibitem{Storo}
A. Storozhenko, P. Schuck, J. Dukelsky, G. Roepke, A. Vdovin, Ann. Physics {\bf 307}, 308 (2003).
\bibitem{Jemai}
M. Jemai, P. Schuck, J. Dukelsky, R. Bennaceur, Phys. Rev. B {\bf 71}, 085115 (2005).
\bibitem{Jemai2}
M. Jemai, D. S. Delion, P. Schuck, Phys. Rev. C {\bf 88}, 044004 (2013).
\bibitem{Bonitz}
M. Bonitz, {\it Quantum kinetic theory, second edition}, Springer 2015.
\bibitem{TS14}
M. Tohyama, P. Schuck, Eur. Phys. J. A {\bf 50}, 77 (2014).
\bibitem{WC}
S. J. Wang and W. Cassing: Ann. Phys. {\bf 159}, 328 (1985).
\bibitem{TS10}
M. Tohyama, P. Schuck, Eur. Phys. J. A {\bf 45}, 257 (2010).
\bibitem{Faddeev}
S. Ethofer, P. Schuck, Z. Physik {\bf 228}, 264 (1969).
\bibitem{RS}
P. Ring and P. Schuck, {\it The nuclear many-body problem}, (Springer-Verlag, Berlin, 1980).
\bibitem{BR}
J.-P. Blaizot, G. Ripka, {\it Quantum Theory of Finite Systems}, The MIT Press, 1986.
\bibitem{TTS04}
M. Tohyama, S. Takahara, P. Schuck, Eur. Phys. J. A {\bf 21}, 217 (2004).
\bibitem{Takahara}
S. Takahara, M. Tohyama, P. Schuck, Phys. Rev. C {\bf 70}, 057307 (2004).
\bibitem{Toh15}
M. Tohyama, Phys. Rev. C {\bf 91}, 017301 (2015).
\bibitem{gell}
M. Gell-Mann and F. Low, Phys. Rev. {\bf 84}, 350 (1951).
\bibitem{lacroix}
M. Assi$\acute{\rm e}$ and D. Lacroix, Phys. Rev. Lett. {\bf 102}, 202501 (2009).
\bibitem{pfitz}
A. Pfitzner, W. Cassing, and A. Peter, Nucl. Phys. A{\bf 577}, 753 (1994).
\bibitem{GT}
M. Gong and M. Tohyama: Z. Phys. A{\bf 335}, 153 (1990).
\bibitem{Rowe}
D. J. Rowe, Rev. Mod. Phys. {\bf 40}, 153 (1968); D. J. Rowe, {\it Nuclear Collective Motion, Models and Theory}, World Scientific 2010.
\bibitem{TS04}
M. Tohyama, P. Schuck, Eur. Phys. J. A {\bf 19}, 215 (2004).
\bibitem{Toh07}
M. Tohyama, Phys. Rev. C {\bf 75}, 044310 (2007).
\bibitem{Kirson}
M. Kirson, Annals of Physics {\bf 66}, 624 (1971).
\bibitem{2ndrpa}
D. Gambacurta, M. Grasso, J. Engel, Phys. Rev. C {\bf 92}, 034303 (2015).
\bibitem{Jansen}
D. Janssen, P. Schuck, Z. Physik A {\bf 339}, 43 (1991).
\bibitem{Cat96}
F. Catara, G. Piccitto, M. Sambataro, N. Van Giai, Phys. Rev. B {\bf 54}, 17536 (1996); F. Catara, M. Grasso, G. Piccitto, M. Sambataro, {\it ibid.} {\bf 58}, 16070 (1998).
\bibitem{Colloquium}
D. S. Delion, P. Schuck, and M. Tohyama, Eur. Phys. J. B {\bf 89}, 45 (2016).
\bibitem{epja32}
M. Tohyama and P. Schuck,  Eur. Phys. J. A {\bf 32}, 139 (2007).
\bibitem{Lipkin3}
D. S. Delion, P. Schuck, J. Dukelsky, Phys. Rev. C {\bf 72}, 064305 (2005).
\bibitem{FF84}
G. Feldman, T. Fulton, Annals of Physics {\bf 152}, 376 (1984).
\bibitem{Lip} 
H. J. Lipkin, N. Meshkov and A. J. Glick, Nucl. Phys. {\bf 62}, 188 (1965).
\bibitem{rich} 
R. W. Richardson, Phys. Rev. {\bf 141}, 949 (1966).
\bibitem{Saar}
M. Saarela, Lecture notes at fall 2008, University of Oulu; Material for reading: A. Fabrocinini, S. Fantoni, E. Krotscheck (Eds0: {\it Introduction to Modern Methods of Quantum Many-Body Theories and their Applications}, Series on Advances in Many Body Theory-Vol. 7, World Scientific, London (2002).
\bibitem{K-Baym}
L. P. Kadanoff, G. Baym, {\it Quantum Statistical Mechanics}, Benjamin, New York, 1962.
\bibitem{pairing-cct1}
T. M. Henderson, G. E. Scuseria, J. Dukelsky, A. Signoracci, T. Duguet, Phys. Rev. C {\bf 89}, 054305 (2014).
\bibitem{pairing-cct2}
M. Degroote, T. M. Henderson, Jinmo Zhao, J. Dukelsky, G. E. Scuseria, arXiv:1512.06111.

\end{thebibliography}
\end{document}